\documentclass[twocolumn,aps,prd,floatfix,superscriptaddress]{revtex4}
\usepackage{latexsym, amsmath, amssymb,amsthm,amsopn,amsfonts, graphicx, epstopdf, multirow}
\usepackage{graphicx}
\usepackage{color}
\usepackage{amssymb}
\usepackage{amsmath}
\usepackage{natbib}
\usepackage{textcomp}
\usepackage{hyperref}
\usepackage{placeins}
\usepackage[dvipsnames]{xcolor}

\DeclareMathOperator{\tr}{tr}

\def\ds{\displaystyle}

\hypersetup{linktoc=all}
\hypersetup{colorlinks = true, allcolors = blue}

\begin{document}

\title{Framework for analysis of next generation, polarised CMB data sets in the presence of galactic foregrounds and systematic effects}

\author{Clara Verg\`es} 
\affiliation{Université de Paris, CNRS, Astroparticule et Cosmologie, F-75006 Paris, France}
\author{Josquin Errard}
\affiliation{Université de Paris, CNRS, Astroparticule et Cosmologie, F-75006 Paris, France}
\author{Radek Stompor}
\affiliation{Université de Paris, CNRS, Astroparticule et Cosmologie, F-75006 Paris, France}
\affiliation{CNRS-UCB International Research Laboratory, ``Centre Pierre Binétruy'', UMI2007, CPB-IN2P3}

\begin{abstract}

Reaching the sufficient sensitivity to detect primordial B-modes requires modern CMB polarisation experiments to rely on new technologies, necessary for the deployment of arrays of thousands of detectors with a broad frequency coverage and operating them for extended periods of time. This increased complexity of experimental design unavoidably introduces new instrumental and systematic effects, which in turn may impact performance of the new instruments. In this work we extend the standard data analysis pipeline by including a (parametric) model of instrumental effects directly in the data model. We then correct for them 
in the analysis, accounting for the additional uncertainty in the final results. We embed these techniques within a general, end-to-end formalism for estimating the impact of the instrument and foreground models on constraints on the amplitude of the primordial B-mode signal. We focus on the parametric component separation approach which we generalize to allow for simultaneous estimation of instrumental and foreground parameters.

We demonstrate the framework by studying in detail the effects induced by an achromatic half-wave plate (HWP), which lead to a frequency-dependent variation of the instrument polarisation angle, and experimental bandpasses which define observational frequency bands. 
We assume a typical Stage-3 CMB polarisation experiment, and show that maps recovered from raw data collected at each frequency band will unavoidably be linear mixtures of the $Q$ and $U$ Stokes parameters. We then derive a new generalized data model appropriate for such cases, and extend the component separation approach to account for it. We find that some of the instrumental parameters, in particularly those describing the HWP can be successfully constrained by the data themselves without need for external information, while others, like bandpasses, need to be known with good precision in advance.

\end{abstract}

\maketitle

\section{Introduction}
\label{intro}

Most of current and future CMB experiments aim at detecting large scale primordial B-modes and measuring the tensor-to-scalar ratio $r$ as low as $r \sim 0.001$. This detection requires unprecedented instrumental sensitivity, calling for  deployment of many thousands of photon-noise limited detectors, along with new technologies at many stages of the detection chain. This increased complexity will introduce new instrumental and systematic effects, which will need to be characterised, accounted for and most likely mitigated. Moreover, the B-modes signal is overshadowed by galactic polarised emission due to thermal dust emission and synchrotron radiation. Separating sky components and isolating the CMB signal is one of the key challenges in the search for cosmological B-modes. This requires the deployment of multi-frequency experiments and multichroic focal planes to take advantage of the fact that different sky components do not have the same frequency scaling.

In this context, it is crucial to develop efficient tools for processing next generation CMB polarisation experiments data sets, to ensure that these experiments achieve the required performance. It is also essential to develop reliable techniques for forecasting the performance of the future experiments, so we can be better prepared for deployment and data analysis. 

CMB data analysis is a complex process composed of many steps. One of its milestones is map-making, that aims at reconstructing single frequency intensity and polarisation maps from raw (time-domain) data collected by a CMB instrument. Another one is component separation, that aims at separating the CMB emission from that of galactic foregrounds.
Component separation methods proposed to date, e.g., \cite{compsep_review,ILC,smica,template-fitting,commander}, typically use maps as produced during the map-making as inputs, and component separation operations are performed in the pixel (or equivalently harmonic) domain. However, many of the instrumental effects can be modelled reliably only in the raw data (time) domain. In this work we therefore consider both steps, map-making and component separation, which we extend by including explicitly models of instrumental systematic effects. We then attempt to correct for them as part of the analysis process.
For definiteness, we focus hereafter on a pixel-based parametric component separation technique~\cite{maxlike, spectral-likelihood} and implement the extensions within the previously validated framework for parametric component separation, {\sc xForecast}~\cite{xforecast,xForecast2}.
In order to demonstrate this general approach, we study in detail the interplay between HWP frequency-dependent characteristics, frequency bandpasses, and galactic foreground emission laws, and their impact on the precision of the primordial $B$-modes detection.  

Frequency-dependent, instrumental effects are particularly insidious for CMB polarisation instruments as they are likely to have significant impact on the component separation procedure, potentially leading to large foreground residuals left in 
the cleaned CMB map. This could bias the derived constraints on the cosmological signal, in particular the tensor-to-scalar ratio $r$ that we seek to measure. We note that while in our worked example we showcase the proposed methodology in the context of performance forecasting, the proposed approach is applicable to actual analysis of the forthcoming CMB data sets.

We derive a new generalised time-domain data model in section~\ref{inst-framework}, where we also discuss its consequences for the map-making procedure and its output. In section~\ref{comp-sep} we elaborate on the parametric component separation procedure and present the self-consistent, end-to-end framework adapted to features of the new data model. We validate and demonstrate the framework in the context of performance forecasting in section~\ref{results}.

\section{Instrumental framework}
\label{inst-framework}

The instrument we consider in this work is composed of a continuously rotating multi-layer half-wave plate and sinuous antennas coupled to total power detectors, such as transition edge sensors (TES) bolometers. In this section, we detail mathematical models describing such an instrument, and use them to derive a corresponding data model of its measurements. We then discuss its consequences for the map-making procedure.

\subsection{Mueller matrix formalism}
\label{Mueller}
The Mueller matrix formalism is widely used to model polarisers in general, and HWP in particular (see for example \cite{HWP-SPIE,LB-HWP}). Incoming light is described by the four Stokes parameters: 

\begin{eqnarray}
\mathbf{I} & \equiv &
\left(
\begin{array}{c}
    I\\
    Q\\
    U\\
    V
\end{array}
\right)
\label{eq:stokes}
\end{eqnarray}
Mueller matrices are $(4 \times 4)$ operators acting on Stokes parameters and describing the impact of different stages of the detection chain on the state of the incoming light.

\subsubsection{Half-wave plate}
\label{hwp}
A single-layer half-wave plate (HWP) is a retarder made of a bi-refringent material that introduces a phase shift of $\pi$ between the input and output polarisation states.
This relationship is true only for a given frequency (monochromatic light). In more generic cases where the light is poly-chromatic, a HWP introduces a phase $\delta$ between polarisation components, which depends on observation frequency. The general Mueller matrix for a single-layer HWP can be written as: 

\begin{eqnarray}
\mathbf{M}_\mathrm{layer} & \equiv &
\left(
\begin{array}{c c c c}
1 & 0 & 0 & 0 \\
0 & 1 & 0 & 0 \\
0 & 0 & \cos\delta & - \sin\delta \\
0 & 0 & \sin\delta &  \phantom{-}\cos\delta 
\end{array}
\right),
\label{eq:one-layer}
\end{eqnarray}
where the phase shift $\delta$ is given by:

\begin{eqnarray}
\delta & \equiv & \frac{2\pi\theta_\mathrm{hwp}\,|n_o-n_e| \nu }{c},
\label{eq:delta}
\end{eqnarray}
where $c$ refers to the usual speed of light in vacuum. Through Eq.~(\ref{eq:delta}) defining $\delta$, coefficients of the HWP Mueller matrix depend on thickness of the bi-refringent layer $\theta_\mathrm{hwp}$, optical indices of the ordinary and extraordinary axes of bi-refringent material $n_o$ and $n_e$, and the frequency of observation $\nu$.

For typical CMB experiments observing in the millimeter range, achromatic HWP are made of several layers of bi-refringent materials, such as sapphire \cite{achromatic-hwp}. Each layer can be modelled as in Eq.~(\ref{eq:one-layer}). 

In this work, we neglect reflections at the interface between two stacked layers, as well as effects of slant incidence angles. We thus model an achromatic HWP as a perfect stack of layers, each layer being rotated with respect to the reference frame of the instrument by an angle $\alpha_i$:

\begin{equation}
    \mathbf{M}_\mathrm{HWP}=\prod_{i = 1}^{ n_\mathrm{layers}}\mathbf{R}(-2\alpha_i)\mathbf{M}_\mathrm{layer, i}\mathbf{R}(2\alpha_i)
    \label{eq:multi-layer-hwp}
\end{equation}
where $\mathbf{R}$ is a rotation matrix.

For the HWP continuously spinning around its axis, on denoting the rotation angle between its axis and the reference frame of the instrument by $\varphi_t$, see Fig.~\ref{fig:angle-def}, we have,
\begin{equation}
   \mathbf{M}_\mathrm{rotating\ HWP}=\mathbf{R}(-2\varphi_t)\mathbf{M}_\mathrm{HWP}\mathbf{R}(2\varphi_t).
    \label{eq:M_layer}
\end{equation}

\subsubsection{Sinuous antennas}
\label{sin-antenna}
The sky signal modulated by the HWP falls then onto an antenna coupled to a detector pair. The antenna separates the orthogonal polarization states of the incident light which are then measured by the detectors. Several future CMB experiments will deploy sinuous antennas, which are broad frequency-band, but introduce a frequency-dependent rotation of the polarization angle of the incident light~\cite{toki-thesis}.

Following~\cite{toki-thesis} we model a sinuous antenna as a usual double-slot antenna (equivalent to a grid), rotated by a frequency-dependent angle $\eta_\nu$ (see Figure \ref{fig:angle-def}), and write its Mueller matrix,
\begin{equation}
    \mathbf{M}_\mathrm{antenna}=\mathbf{R}(-2 \eta_\nu) \mathbf{M}_\mathrm{grid} \mathbf{R}(2 \eta_\nu)
\end{equation}
where~\cite{toki-thesis}
\begin{equation}
    \eta_\nu\,[\deg]\,\simeq\, 4.9 \, \sin \left(12 \, \log(\nu) + 4.7\right)
\end{equation}
and
\begin{eqnarray}
\mathbf{M}_\mathrm{grid} & \equiv &
\frac{1}{2}\left(
\begin{array}{c c c c}\medskip
1 & 1 & 0 & 0 \\\medskip
1 & 1 & 0 & 0 \\\medskip
0 & 0 & 0 & 0\\\medskip
0 & 0 & 0 & 0
\end{array}
\right).
 \end{eqnarray}
The complete Mueller matrix of this optics system can therefore be written as,
\begin{equation}
    \mathbf{M}_\mathrm{optics}=\mathbf{M}_\mathrm{antenna}\mathbf{R}(-2\varphi_t)\mathbf{M}_\mathrm{HWP}\mathbf{R}(2\varphi_t).
    \label{eq:M_optics}
\end{equation}
This matrix is the transfer function of the optical system in the Stokes parameter domain, applied to the incoming light represented by the four Stokes parameters, Eq.~(\ref{eq:stokes}). In what follows, for definiteness, we assume that circular polarization of the incident light is negligible and we set it to zero i.e., $\mathrm{V}_\mathrm{in} = 0$. This is consistent with the expectation that for the relevant cosmological, astrophysical, or atmospheric signals, $V$ is either zero or  negligible~\cite{vpolar-atm,sync-vpolar}.

\subsection{Time domain data model}
\label{time-domain}
\begin{figure}
    \centering
    \includegraphics[width=0.5\textwidth]{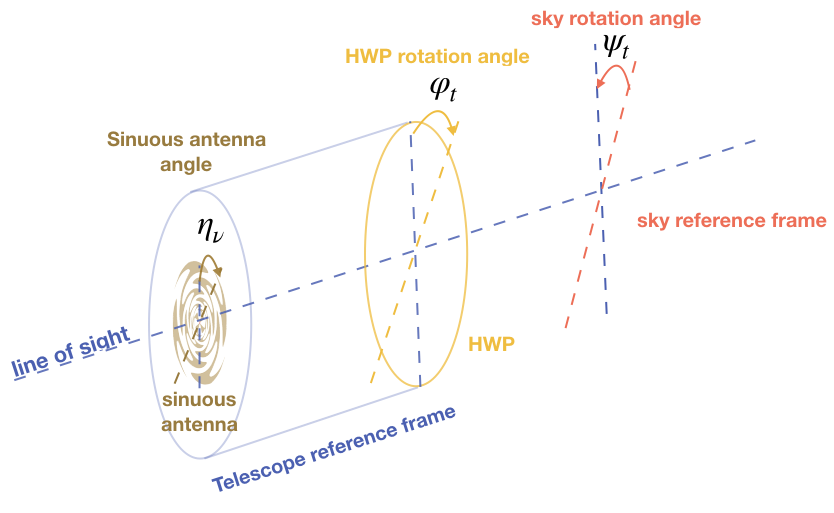}
    \caption{Depiction of sky rotation angle, $\psi_t$, defined as an angle between the sky reference frame and the telescope reference frame and of HWP rotation, $\varphi_t,$ and sinuous antenna angles, $\eta_\nu$, both assumed to be measured with respect to the telescope reference frame.}
    \label{fig:angle-def}
\end{figure}

During observations, CMB telescopes scan the sky and their orientation with respect to the sky changes. This introduces a time-dependent rotation angle, $\psi_t$, defining the position of the instrument with respect to sky coordinates (see Figure \ref{fig:angle-def}), modifying the Mueller matrix:
\begin{equation}
    \mathbf{M}_\mathrm{tot} = \mathbf{M}_\mathrm{optics} \mathbf{R}(2 \psi_t).
    \label{eq:M_tot}
\end{equation}

\subsubsection{Single layer HWP}
\label{1layer-hwp}
In the simple case of a monochromatic single layer rotating HWP with grid antenna, the Mueller matrix of the optics system reads,

\begin{equation}
    \mathbf{M}_\mathrm{optics, mono.} = \mathbf{M}_\mathrm{grid} \mathbf{R}(-2\varphi_t) \mathbf{M}_\mathrm{layer}(\delta = \pi)\mathbf{R}(2\varphi_t),
\end{equation}
what is a special case of Eq.~(\ref{eq:M_optics}). Taking into account sky rotation as in Eq.~(\ref{eq:M_tot}), the time domain data model for an instrument pointing in the direction $\gamma$ at time $t$, can be written as:
\begin{align}
    \mathbf{m}_t(\nu) & =  \mathrm{I}(\gamma_t, \nu) \nonumber \\
    & + \cos(4\varphi_t+2\psi_t)\,\mathrm{Q}(\gamma_t, \nu) \nonumber \\
     & + \sin(4\varphi_t+2\psi_t)\,\mathrm{U}(\gamma_t, \nu) \nonumber \\
     & + \, \mathbf{n}_t.
     \label{eq:basicPolDataModel}
\end{align}

This is the standard data model of a HWP-modulated CMB measurement, as assumed in the standard map-making procedure. While all three Stokes parameter maps are mixed together in every single measurement, we can separate them from each other owing to a different time dependence of their respective coefficients: the signals modulated in Eq.~(\ref{eq:basicPolDataModel}) as $\cos(4\varphi_t+2\psi_t)$ are those of the $Q$ parameter, those modulated as $\sin(4\varphi_t+2\psi_t)$ of $U$ and the non-modulated ones of the total intensity. This is done during the map-making procedure which thus produces three sky maps of $I$, $Q$ and $U$ for each frequency band.

\subsubsection{Multi layer HWP}
\label{multilayer-hwp}
In a more general case of the multi-layer HWP, the detected signal can be expressed as:
\begin{align}
\label{eq:simple-model}
    \mathbf{m}_t(\nu) & =  \mathbf{M}_{00}(\nu)\,\mathrm{I}(\gamma_t, \nu) \nonumber \\
    & + \, \mathbf{M}_{01}(\nu, \varphi_t,\psi_t)\,\mathrm{Q}(\gamma_t, \nu) \nonumber \\ 
    & + \mathbf{M}_{02}(\nu,\varphi_t,\psi_t)\,\mathrm{U}(\gamma_t, \nu) \\ \nonumber 
    & + \, \mathbf{n}_t
\end{align}
where $\mathbf{M}_{0\mathrm{i}}$ denotes the elements of the first row of $\mathbf{M}_{\rm tot}$, Eq.~(\ref{eq:M_tot}). These elements can be represented as:
\begin{align}
\mathbf{M}_{0i}(\varphi_t, \psi_t; \nu, \nu) & = 
\sum_{k = 0, 4}\,
\mathbf{C}_{0i;\, k}(\nu)\,\cos(k\varphi_t+2\psi_t) \nonumber\\
& +  \sum_{k = 0, 4}\,
\mathbf{S}_{0i;\, k}(\nu)\,\sin(k\varphi_t+2\psi_t), 
\label{eq:decomposition}
\end{align}
where we have introduced $\mathbf{C}_{0i;\, k}$ (resp. $\mathbf{S}_{0i;\, k}$), the coefficients of the cosine (resp. sine) modulated terms, which are linear combinations of elements of the Mueller matrix of the system, $\mathbf{M}_{0\mathrm{i}}$. They therefore depend explicitly on instrumental, i.e., here HWP and sinuous antennas parameters, as well as observational frequency. We give explicit expressions for these coefficients in Appendix~\ref{hwp-appendix}.

Combining Eqs.~(\ref{eq:simple-model}) and (\ref{eq:decomposition}) we can rewrite the data model by grouping together terms with the same time dependence:
\begin{widetext}
  \begin{align}
\mathbf{m}_t(\nu) & \equiv
\mathbf{n}_t  \, + \, \mathbf{M}_{00}(\nu)\,\mathrm{I}(\gamma_t, \nu) \nonumber\\
& + \Big[\,
\mathbf{C}_{01;\, 0}(\nu)\,\mathrm{Q}(\gamma_t, \nu)\,+\,
\mathbf{C}_{02;\, 0}(\nu)\,\mathrm{U}(\gamma_t, \nu)\;\Big] \, \times \cos 2\psi_t\nonumber\\
& +\Big[\,
\mathbf{S}_{01;\, 0}(\nu)\,\mathrm{Q}(\gamma_t, \nu)\, + \,
\mathbf{S}_{02;\, 0}(\nu)\,\mathrm{U}(\gamma_t, \nu)\; \Big] \, \times \sin 2\psi_t\nonumber\\
& + 
\Big[ \, 
\mathbf{C}_{01;\, 4}(\nu)\,\mathrm{Q}(\gamma_t, \nu)\,
+ \,\mathbf{C}_{02;\, 4}(\nu)\,\mathrm{U}(\gamma_t, \nu) \;\Big]
 \times \, \cos (4\phi_t+2\psi_t)
\nonumber\\
& + \Big[\, \mathbf{S}_{01;\, 4}(\nu) \,\mathrm{Q}(\gamma_t, \nu)\,+ \,\mathbf{S}_{02;\, 4}(\nu)\,\mathrm{U}(\gamma_t, \nu)\,\Big]\,\times \, \sin(4\phi_t+2\psi_t).
\label{eq:datamodel-simple}
\end{align}
\end{widetext}
This expression highlights the fact that in the case of a multi-layer HWP, Stokes parameters $Q$ and $U$ of the sky signal are not simply modulated at $\cos(4\varphi_t+2\psi_t)$ $\sin(4\varphi_t+2\psi_t)$, respectively. Instead, and in contrast to the case of the simple polarized data model in Eq.~(\ref{eq:basicPolDataModel}), the terms modulated in this way are composed of mixtures of these two Stokes parameters. We refer to those signals hereafter as \emph{mixed-Stokes single frequency signals}. We also note the presence of two extra terms which are modulated by the sky angle only. Sky signals, they correspond to, constitute two additional, independent combinations of the Stokes $Q$ and $U$ parameters. As the coefficients $\mathbf{C}_{0\mathrm{i};\, \mathrm{k}}$ and $\mathbf{S}_{0\mathrm{i};\, \mathrm{k}}$, depend on instrumental parameters (see Appendix~\ref{hwp-appendix}) so do the mixed-Stokes signals, which are therefore specific to a given experiment.

As mentioned earlier, the map-making procedure enables an estimation of the terms with different time dependence, assuming that those dependencies are such that they are linearly independent when limited to observations of a single sky pixel for every such a pixel. If these conditions are met, the map-making procedure applied to the data model in Eq.~(\ref{eq:datamodel-simple}) could recover for each frequency channel a map of total intensity accompanied by four maps composed of different linear combinations of Stokes $Q$ and $U$ parameters. These would be those maps which could and should be considered as inputs to the next data processing stages and specifically that of the component separation. We explore some of the consequences of this fact in the reminder of this article.

One important implication of the more involved data model in Eq.~(\ref{eq:datamodel-simple}) is that solving the complete map-making problem in this case will in general require sufficient redundancy in observations of every sky pixel, with a number of different HWP as well as sky rotation angles. This can have important impact on scan designs of the future CMB experiments. 

In our case we will however assume a perfectly uniform sky coverage, both in terms of the number of observations of each pixel, as well as distributions of the HWP and sky angles. With these assumptions the map-making problem for the data model in Eq.~(\ref{eq:datamodel-simple}) is not only solvable, but results in estimates of the five sky signals which have mutually uncorrelated noise. Assuming furthermore that the instrumental noise  is white, the RMS of the noise in the four mixed-Stokes maps and one total intensity is the same as in the standard
case, i.e.,
\begin{eqnarray}
\sigma_{RMS}(\mathrm{POL}) \; = \; \sigma_{RMS}(\mathrm{INT})\,\sqrt{2} \; \propto \; \sqrt{\frac{2}{n_\mathrm{hits}}}
\end{eqnarray}
In the following, we use these estimates as reflecting the noise levels in our mixed-Stokes maps.

For shortness, hereafter we denote the four mixed-Stokes signals as $\boldsymbol{\mathcal{C}}_0$, $\boldsymbol{\mathcal{S}}_0$, $\boldsymbol{\mathcal{C}}_4$, $\boldsymbol{\mathcal{S}}_4$, and introduce an effective intensity signal, $\boldsymbol{\mathcal{I}}(\gamma_t, \nu) \equiv \mathbf{M}_{00} \mathrm{I}(\gamma_t, \nu)$, rewriting our data model as,
\begin{align}
    \mathbf{m}_t(\nu) & \equiv \boldsymbol{\mathcal{I}}(\gamma_t, \nu) \nonumber \\ 
    & + \boldsymbol{\mathcal{C}}_0(\gamma_t, \nu) \cos (2\psi_t) 
    \nonumber\\ & 
+ \boldsymbol{\mathcal{S}}_0(\gamma_t, \nu) \sin(2\psi_t) \nonumber\\
& + \boldsymbol{\mathcal{C}}_4(\gamma_t, \nu) \cos (4\phi_t+2\psi_t)
\nonumber\\ & 
+ \boldsymbol{\mathcal{S}}_4(\gamma_t, \nu) \sin(4\phi_t+2\psi_t) \nonumber \\
& + \mathbf{n}_t.
\label{eq:generalised_stokes_def}
\end{align}
These mixed Stokes components are related to standard Stokes components, see Eq.~(\ref{eq:datamodel-simple}), via a linear transformation given by:

\begin{eqnarray}
\left(
\begin{array}{c}
\boldsymbol{\mathcal{I}}(\gamma_t, \nu)\\
\boldsymbol{\mathcal{C}}_0(\gamma_t, \nu)\\
\boldsymbol{\mathcal{S}}_0(\gamma_t, \nu)\\
\boldsymbol{\mathcal{C}}_4(\gamma_t, \nu)\\
\boldsymbol{\mathcal{S}}_4(\gamma_t, \nu)
\end{array}
\right) & = & \underbrace{\left[
\begin{array}{ccc}
 \mathbf{M}_{00}(\nu) & 0 & 0\\
 0 & \mathbf{C}_{01;0}( \nu) &
\mathbf{C}_{02;0}( \nu)
\\
0 & \mathbf{S}_{01;0}( \nu) &
\mathbf{S}_{02;0}( \nu)
\\
0 & \mathbf{C}_{01;4}( \nu) &
\mathbf{C}_{02;4}( \nu)
\\
0 & \mathbf{S}_{01;4}( \nu) &
\mathbf{S}_{02;4}( \nu)
\end{array}\right]}_{\ds \equiv \mathbf{M}(\nu)}\nonumber \\ 
&\times& \left(
\begin{array}{c}
    \mathrm{I}(\gamma_t, \nu)\\
    \mathrm{Q}(\gamma_t, \nu)\\
    \mathrm{U}(\gamma_t, \nu)\\
\end{array}
\right), 
\label{eq:data-model-no-bp}
\end{eqnarray}
where the transformation matrix, $\mathbf{M}(\nu)$, depends on instrumental parameters. We conclude therefore that the map-making procedure could render estimates of pure $Q$ and $U$ Stokes parameters only if the true values of the relevant instrumental parameters are known.

Lastly, we note that the fact that the mixed Stokes signals are combinations of only Stokes Q and U parameters is not universal. Indeed, this is, for instance, not the case whenever I-to-P leakage is present. The formalism presented hereafter can be straightforwardly extended to include such cases.

\subsubsection{Bandpasses}
\label{bandpasses}
The actual CMB experiments detect the incoming signal integrated over some frequency bands rather than the monochromatic one. The frequency bands are centered at desired frequencies, $\nu_c$, and are defined by the instrumental response functions referred to hereafter as bandpasses. These are denoted as $\mathcal{B}(\nu, \nu_c)$. To account for their effects we need to introduce bandpass-averaged quantities, which we distinguish with a bar over a symbol and defined as:

\begin{equation}
    {\bar{\mathbf{X}}}(\nu_c) \; \equiv \; \frac{\int d\nu \, \mathcal{B}(\nu, \nu_c)\, \mathbf{X}(\nu)\, k(\nu)}{\int d\nu \, \mathcal{B}(\nu) \, k(\nu)},
    \label{eq:integrated_x}
\end{equation}
where $k$ is a conversion factor which reconciles the units used for the data, $\mathbf{X}$, and the bandpasses, $\mathcal{B}$.

We can then rewrite the data model in Eq.~(\ref{eq:generalised_stokes_def}) using the bandpass averaged objects as:
\begin{align}
    \bar{\mathbf{m}}_t(\nu_c) & \equiv \boldsymbol{\bar{\mathcal{I}}}(\gamma_t, \nu_c) \nonumber \\ 
    & + \boldsymbol{\bar{\mathcal{C}}}_0(\gamma_t, \nu_c) \cos (2\psi_t) 
    \nonumber\\ & 
+ \boldsymbol{\bar{\mathcal{S}}}_0(\gamma_t, \nu_c) \sin(2\psi_t) \nonumber\\
& + \boldsymbol{\bar{\mathcal{C}}}_4(\gamma_t, \nu_c) \cos (4\phi_t+2\psi_t)
\nonumber\\ & 
+ \boldsymbol{\bar{\mathcal{S}}}_4(\gamma_t, \nu_c) \sin(4\phi_t+2\psi_t) \nonumber \\
& + \bar{\mathbf{n}}_t.
\label{eq:data-model-w-bp}
\end{align}
The map-making codes can be used assuming this data model in the same way as in the monochromatic case however they will now produce bandpass-averaged maps of mixed-Stokes and effective total intensity.

We further note that going from these maps to the maps of pure Stokes parameters for each frequency band is not possible once the bandpasses are explicitly included. This is because the integration over the bandpasses does not preserve the matrix form of Eq.~(\ref{eq:data-model-no-bp}) as the pure Stokes signals, $I(\gamma_t, \nu)$, $Q(\gamma_t, \nu)$, $U(\gamma_t, \nu)$, are integrated over the frequency together with the corresponding elements of the matrix $\mathbf{M}(\nu)$. Consequently, the mixed-Stokes maps are the only available objects at the end of the map-making even if the instrumental parameters were perfectly known. This does not however prevent us from recovering pure Stokes parameters of maps of components of a different physical origin. We discuss this in the following, 

\subsection{Multi-component data model}
\label{multi-comp}
The polarized sky signal as measured at different frequencies and characterised by three Stokes parameters, $I$, $Q$, $U$, is really a mixture of the genuine CMB signal and polarised galactic foregrounds. The two most important polarised  foregrounds are polarised thermal dust emission and synchrotron radiation. Separating the measured signals into signals of the sky components is the goal of component separation procedures. Those typically are performed in the pixel domain and use maps produced on the map-making step as inputs. In this section we derive a data model for the mixed-Stokes maps, Eq.~(\ref{eq:data-model-w-bp}), which in the discussed case constitute the output of the map-making procedure and therefore the input of the component separation.

\subsubsection{Sky components}
We model foregrounds by a sky template at a reference frequency $\nu_0$, scaled to the observation frequency $\nu$ using foregrounds emission models. The effective signal as measured at each frequency is a mixture of these three major sky components (CMB, dust and synchrotron), which is described by a mixing matrix $\mathbf{A}$ containing the scaling laws for each sky component at each frequency.
The scaling laws we assume in this work are fairly standard, e.g., \cite{xForecast2}, and in the $\mu$K$_\mathrm{RJ}$ units they are given by:
\begin{equation}
    \mathbf{A}^\mathrm{sync}(\nu, \nu_0) = \bigg( \frac{\nu}{\nu_0}\bigg)^{\beta_s}
\end{equation}
where $\beta_s$ is the synchrotron spectral index, and 

\begin{equation}
    \mathbf{A}^\mathrm{dust}(\nu,\nu_0) = \bigg( \frac{\nu}{\nu_0}\bigg)^{\beta_d+1}\,\frac{\exp{\frac{h\nu_0}{kT_d}}-1}{\exp{\frac{h\nu}{k T_d}}-1}
\end{equation}
where $\beta_d$ is the dust spectral index and $T_d$ is the dust temperature.

The CMB follows a known black body emission spectrum, with $T_\mathrm{CMB} = 2.7255 \, K$, leading to the assumption that $\mathbf{A}^\mathrm{CMB}(\nu,\nu_0) = 1$ in $\mu$K$_\mathrm{CMB}$ units.

For a given observed frequency $\nu$, the sky signal can be modeled as:

\begin{align}
\label{eq:mixing_matrix}
    \left(
\begin{array}{c}
    \mathrm{I}(\nu)\\
    \mathrm{Q}(\nu)\\
    \mathrm{U}(\nu)\\
\end{array}
\right) 
= \sum_{\genfrac{}{}{0pt}{2}{\mathrm{comp}=\mathrm{cmb}}{\mathrm{dust},\mathrm{sync}}} \mathbf{A}^\mathrm{comp}(\nu, \nu_0)    \left(
\begin{array}{c}
    \mathrm{I}_\mathrm{comp}(\nu_0)\\
    \mathrm{Q}_\mathrm{comp}(\nu_0)\\
    \mathrm{U}_\mathrm{comp}(\nu_0)\\
\end{array}
\right),
\end{align}
with $\mathbf{A}^\mathrm{comp}$ a diagonal matrix  scaling each component template to the desired frequency, $\nu$.  The latter, as presented here, are in $\mu$K$_\mathrm{RJ}$, but for simplicity, we generate maps at all frequencies and for all components in $\mu$K$_\mathrm{CMB}$ units, as are the recovered maps of the CMB and foreground signals. A conversion factor is then taken into account so that units are handled consistently throughout the process.

\subsubsection{Multi-component sky model}
To derive a complete multi-component model of the mixed-Stokes maps, we first introduce explicitly the different sky components in the model in Eq.~(\ref{eq:data-model-no-bp}) using Eq.~(\ref{eq:mixing_matrix}):

\begin{widetext}

\begin{align}
\underbrace{\left(
\begin{array}{c}
\mathcal{I}(\gamma_t, \nu)\\
\mathcal{C}_0(\gamma_t, \nu)\\
\mathcal{S}_0(\gamma_t, \nu)\\
\mathcal{C}_4(\gamma_t, \nu)\\
\mathcal{S}_4(\gamma_t, \nu)
\end{array}
\right)} 
_{\ds \equiv \, \mathbf{s}(\gamma_t, \nu)}
& \equiv
\sum_{\genfrac{}{}{0pt}{2}{\mathrm{comp}=\mathrm{cmb}}{\mathrm{dust},\mathrm{sync}}} \;
\underbrace{
\left[
\begin{array}{ccc}
 \mathbf{M}_{00;\, \cos_0}(\nu) & 0 & 0\\
 0 & \mathbf{C}_{01;\, 0}( \nu) &
\mathbf{C}_{02;\, 0}( \nu)
\\
0 & \mathbf{S}_{01;\, 0}( \nu) &
\mathbf{S}_{02;\, 0}( \nu)
\\
0 & \mathbf{C}_{01;\, 4}( \nu) &
\mathbf{C}_{02;\, 4}( \nu)
\\
0 & \mathbf{S}_{01;\, 4}( \nu) &
\mathbf{S}_{02;\, 4}( \nu)
\end{array}\right] \, 
 \mathbf{A}^\mathrm{comp}(\nu,\nu_0)}_{\ds \equiv\, \boldsymbol{\mathcal{A}}^\mathrm{comp}( \nu, \nu_0)}\;
\underbrace{\left(
\begin{array}{c}
\mathrm{I}_\mathrm{comp}(\gamma_t, \nu_0)     \\
\mathrm{Q}_\mathrm{comp}(\gamma_t, \nu_0)
\\
 \mathrm{U}_\mathrm{comp}(\gamma_t, \nu_0)
\end{array}
\right)}_{\displaystyle\equiv \, \mathbf{c}_\mathrm{comp}(\gamma_t, \nu_0)}.
 \label{eq:model-with-comp}
\end{align}

\end{widetext}
We then average both sides of this equation over the bandpasses using the prescription in Eq.~(\ref{eq:integrated_x}), and define component-specific, bandpass-integrated matrices, $\boldsymbol{\mathcal{\bar{A}}}$ as:

\begin{equation}
    \boldsymbol{\mathcal{\bar{A}}}^\mathrm{comp}(\nu_c, \nu_0) \equiv \frac{\int d\nu \, \mathbf{M}(\nu) \, \mathcal{B}(\nu, \nu_c)\, \mathbf{A}^\mathrm{comp}(\nu, \nu_0) \, k(\nu)}{\int d\nu \, \mathcal{B}(\nu) \, k(\nu)}
    \label{eq:integrated_m},
\end{equation}
where we assumed that the bandpass center frequency is $\nu_c$.

We can now finally write our multi-component, bandpass-integrated data model for the mixed-Stokes maps as:

\begin{widetext}
\begin{align}
\mathbf{\bar{s}}(\gamma_t, \nu_c,\nu_0)
& = 
\underbrace{\left[
\begin{array}{c c c}
\boldsymbol{\mathcal{\bar{A}}}^\mathrm{cmb}(\nu_c,\nu_0), &  \boldsymbol{\mathcal{\bar{A}}}^\mathrm{dust}(\nu_c,\nu_0), & \boldsymbol{\mathcal{\bar{A}}}^\mathrm{sync}(\nu_c,\nu_0)
\end{array}
\right]}_{\ds \equiv \boldsymbol{\mathcal{\bar{A}}}(\nu_c,\nu_0)}\;
\underbrace{\left(
\begin{array}{c}
\mathbf{c}_\mathrm{cmb}(\gamma_t, \nu_0)     \\
\mathbf{c}_\mathrm{dust}(\gamma_t, \nu_0)
\\
\mathbf{c}_\mathrm{sync}(\gamma_t, \nu_0)
\end{array}
\right)}_{\ds \equiv \mathbf{c}(\gamma_t, \nu_0)} \nonumber \\
& = \boldsymbol{\mathcal{\bar{A}}}(\nu_c,\nu_0) \, \mathbf{c}(\gamma_t, \nu_0).\label{eq:genMixingModel}
\end{align}
\end{widetext}

We point out that while this equation is more involved than the usual relation between the standard observables (i.e., single frequency maps of each Stokes parameter) and the underlying components, c.f., Eq.~(\ref{eq:mixing_matrix}), the relation in both these cases is linear as far as sky component amplitudes are concerned. However the system (mixing) matrix in the mixed-Stokes case depends on both foreground spectral parameters as well as the instrumental parameters.

We explore consequences of these two observations in the follow-up sections.

\begin{figure}
    \centering
    \includegraphics[width=0.5\textwidth]{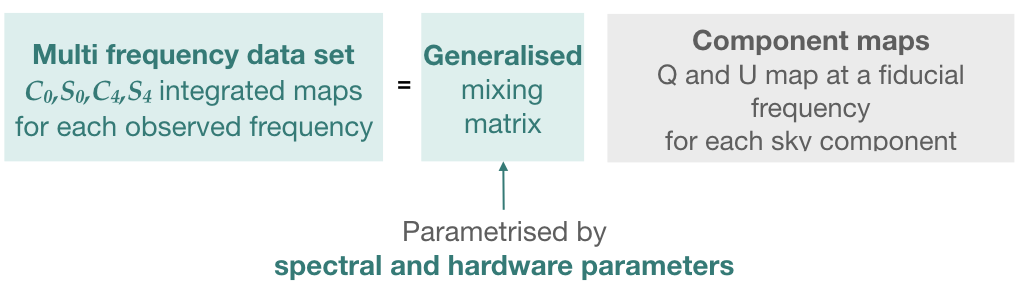}
    \caption{Data model with generalised mixing matrix, Eq.~(\ref{eq:genMixingModel}).}
    \label{fig:data-model}
\end{figure}

\section{Component separation}
\label{comp-sep}

In this section, we follow the formalism of \cite{xforecast}, adapting their procedure to account for mixed Stokes components to include instrumental parameters (HWP, sinuous antenna, bandpasses). We first elaborate the mixed component maps data model,
(\ref{data}) and subsequently discuss component separation and estimation of foreground and hardware parameters from these mixed-Stokes maps (\ref{parameter-est}). With the estimated parameters, we reconstruct the sky signal and estimate residuals (\ref{res}), as well as noise (\ref{noise}) after component separation. Finally, we evaluate the impact of the studied effects on cosmological parameters (\ref{cosmo-likelihood}).
\subsection{Data model}
\label{data}
Collecting together all single-frequency mixed-Stokes maps, we can write our data model for the entire multi frequency data set:
\begin{equation}
    \mathbf{\bar{m}} = \mathbf{\bar{s}} + \mathbf{\bar{n}} \equiv \boldsymbol{\mathcal{\bar{A}}} \, \mathbf{c} + \mathbf{\bar{n}},
\end{equation}
Here the combined data vector $\mathbf{\bar{m}}$ includes all the bandpass-integrated mixed-Stokes maps, $\boldsymbol{\bar{\mathcal{C}}_0}$, $\boldsymbol{\bar{\mathcal{C}}_4}$, $\boldsymbol{\bar{\mathcal{S}}_0}$, $\boldsymbol{\bar{\mathcal{S}}_4}$, as measured for all frequency bands which are all concatenated together in a single data vector. From now on we focus only on the polarisation and exclude total intensity from our consideration. While this is not the most general case, the generalisation to all three Stokes parameters is straightforward.
$\mathbf{\bar{s}}$ is the noiseless signal in the all mixed-Stokes maps recovered at all frequencies, and $\boldsymbol{\mathcal{\bar{A}}}$ is a generalised, multi-frequency mixing matrix. It is composed of single channel matrices, $\boldsymbol{\mathcal{\bar{A}}}(\nu_c, \nu_0)$ defined in Eq.~(\ref{eq:genMixingModel}), put on top of each other.
$\mathbf{\bar{n}}$ denotes the actual noise present in all the mixed-Stokes maps as derived by the map-making procedure.

To improve readability, as we always integrate over bandpass hereafter, we drop the bar that indicates bandpass integration in the follow-up.

For a given pixel $p$, we have:
\begin{equation}
        \mathbf{m}_p = \mathbf{s}_p + \mathbf{n}_p \equiv \boldsymbol{\mathcal{A}}_p(\beta_f,\beta_h)\, \, \mathbf{c}_p + \mathbf{n}_p
        \label{eq:data-model-pixel}
\end{equation}
where $\beta_{f}$ refers to foregrounds spectral parameters and $\beta_{h}$ to hardware parameters. Whereas $\beta_f$ parameters apply to a specific component (dust or synchrotron) and can vary between sky pixels, $\beta_h$ are typically global parameters, applied to all pixels and all sky components --- including CMB. For simplicity, we hereafter assume that the foreground parameters $\beta_f$ that we fit for are not pixel-dependent (e.g., we consider a single set of $\beta_f$ for the entire observed sky). The generalisation of the proposed formalism to pixel-dependent foreground parameters is possible following the steps already outlined in~\cite{xForecast2}.

We also note that the proposed formalism can be extended to allow for the variability of some of the detector properties across the detector arrays thus permitting to study their impact on the experiment performance. 
Such arrays are envisaged for the future generation of the CMB experiments, making this feature of the framework very timely. We leave detailed investigation of such issues to future work and in the following we give an example of the impact of a mismatch between the parameters used for generating the data and those assumed in their modeling in the case of bandpasses. The mismatch could arise in practice as a difference between effective bandpass parameters resulting from averaging data of many detectors and a single set of idealized parameters assumed in the reconstruction.

\subsection{Parameter estimation}
\label{parameter-est}
The generalised data model, although more involved than the standard model assumed by component separation methods, retains all the essential features of the latter and most of the existing component separation procedures can be adapted to account for the extensions. In this work, we focus on the parametric component separation technique as a method of choice for the component separation and in particular its two step implementation as proposed in  \cite{spectral-likelihood} and elaborated on in \cite{ xforecast,xForecast2,errard2011}. We discuss below essential ingredients of the method, emphasising the new features due to the specificity of the new data model.

\subsubsection{Ensemble average likelihood}
\label{likelihood}
For the data model in Eq.~(\ref{eq:data-model-pixel}), we write an effective spectral likelihood~\cite{spectral-likelihood}.
This is obtained by replacing the standard mixing matrix $\mathbf{A}$ (see, e.g.,  Eq.~(6) of~\cite{xforecast}) by the generalised mixing matrix, $\boldsymbol{\mathcal{A}}$. This generalised log-likelihood reads therefore as:

\begin{equation}
    \mathcal{S} = - \sum_p (\boldsymbol{\mathcal{A}}_p^t \, \mathbf{N}_p^{ -1} \, \mathbf{m}_p)^t \, (\boldsymbol{\mathcal{A}}_p^t \, \mathbf{N}_p^{-1} \, \boldsymbol{\mathcal{A}}_p)^{-1}\boldsymbol{\mathcal{A}}_p^t \, \mathbf{N}_p^{ -1} \mathbf{m}_p
    \label{spec_likelihood}
\end{equation}
where $\mathbf{N}_p$ is the noise covariance matrix. A key difference with the standard case is that now the mixing matrix, $\boldsymbol{\mathcal{A}}$, depends on both foreground and instrumental parameters. While ideally the latter are known with sufficient precision from an instrument calibration campaign, this is hardly the case for most of the parameters of interest. Here, we therefore aim at determining both these sets of parameters, i.e., $\beta_f$ and $\beta_h$, from the available data set via the maximisation of the spectral likelihood. Clearly, this may not be always possible and some external information may be required or beneficial, as detailed in section~\ref{priors} and \ref{with_priors}.
This procedure can be applied to any specific input data set, thus providing a basis for actual data processing framework. In the context of performance forecasting we are interested in quantities averaged over statistical ensembles of the possible input data sets. In view of this, we derive ensemble averaged version of the likelihood in Eq.~(\ref{spec_likelihood})~\cite{xforecast}:

\begin{equation}
    \langle \mathcal{S} \rangle = - \tr \, \sum_{p} \, \lbrace (\mathbf{N}^{-1}_p \, - \, \mathbf{P}_p) \, (\langle \mathbf{s}_p\,\mathbf{s}_p^t\rangle \, + \,  \mathbf{N}_p) \rbrace
    \label{eq:S_no_cmb}
\end{equation}
where $\mathbf{P}_p$ is the projection operator defined as:
\begin{equation}
    \mathbf{P}_p \equiv \mathbf{N}^{-1}_p \, - \, \mathbf{N}^{-1}_p \, \boldsymbol{\mathcal{A}_p} ( \boldsymbol{\mathcal{A}_p}^t \, \mathbf{N}^{-1}_p \, \boldsymbol{\mathcal{A}})_p^{-1} \, \boldsymbol{\mathcal{A}}_p^t \,  \mathbf{N}^{-1}_p.
\end{equation}
The single-pixel data covariance matrix in Eq.~(\ref{eq:S_no_cmb}), $\langle \mathbf{s}_p\,\mathbf{s}_p^t \rangle$, reflects the properties of what we assume to be the true sky. Unlike in the case studied in~\cite{xforecast} it has to now explicitly account for the presence of the CMB contribution. This is due to the fact that instrumental parameters can, and do, affect the frequency scaling of the CMB signal.
In order to calculate the covariance we split $\boldsymbol{\mathcal{A}}_p$ and $\mathbf{c}_p$ into CMB and foregrounds parts:

\begin{align}
\mathbf{c}_p & \equiv 
\left[
\begin{array}{l}\medskip
\mathbf{c}_p^\mathrm{cmb}\\
\mathbf{c}_p^\mathrm{fg}
\end{array}
\right]
\label{eq:sDef}
\\
\boldsymbol{\mathcal{A}} & =  
\left[ \boldsymbol{\mathcal{A}}_p^\mathrm{cmb}, \boldsymbol{\mathcal{A}}_p^\mathrm{fg}\right]
\end{align}
and we split the single-pixel data covariance matrix into its CMB and foregrounds contributions accordingly:
\begin{align}
\medskip
\langle \mathbf{s}_p\mathbf{s}_p^t\rangle
& =  \boldsymbol{\mathcal{A}}_p^\mathrm{cmb}\langle \mathbf{c}_p^\mathrm{cmb} \mathbf{c}_p^{\mathrm{cmb},t} \rangle \boldsymbol{\mathcal{A}}_p^{\mathrm{cmb},t} +
\boldsymbol{\mathcal{A}}_p^\mathrm{fg}\mathbf{c}_p^\mathrm{fg}\mathbf{c}_p^{\mathrm{fg}, t} \boldsymbol{\mathcal{A}}_p^{\mathrm{fg},t}
\label{eq:data-covariance-matrix}
\end{align}
The foregrounds contribution (the second term) is easy to compute as we consider foregrounds as  a fixed template independent on the realisation. This is not the case for the CMB contribution though, and in principle the average in the first term of the above equation should include the pixel-pixel correlations. However, if we assume that neither the pointing matrices, $\boldsymbol{\mathcal{A}}_p$, nor the noise, $\mathbf{N}_p$, are  pixel-dependent and that the noise is white, we do not need the full multi-pixel covariance matrix for the CMB but only its single pixel version.
Considering the two polarisation states $Q$ and $U$,

\begin{equation}
    \mathbf{c}^\mathrm{cmb}_p \equiv [ \mathbf{c}^\mathrm{cmb}_{p,Q},\mathbf{c}^\mathrm{cmb}_{p,U}],
\end{equation}
the single-pixel covariance of the CMB signal is a $2$-by-$2$ matrix, $\mathbf{\hat S}^\mathrm{cmb}\equiv \langle \mathbf{c}_p^\mathrm{cmb} \mathbf{c}_p^{\mathrm{cmb},t} \rangle$, given by:
\begin{align}
\mathbf{\hat S}^\mathrm{cmb} 
& \equiv \left[
\begin{array}{cc}\medskip
\mathbf{\sigma}^2_{QQ} & \mathbf{\sigma}^2_{QU} \\
\mathbf{\sigma}^2_{UQ} & \mathbf{\sigma}^2_{UU}
\end{array}
\right],
\label{eq:cmb-data}
\end{align}
and can be straightforwardly calculated for any standard theoretical model.

We can write the total ensemble average likelihood explicitly as:
\begin{equation}
    \mathcal{S} =  - \tr \Big\{ n_\mathrm{pix} \,(\mathbf{N}^{-1} - \mathbf{P}) (\mathbf{F} +\mathbf{N} + \boldsymbol{\mathcal{A}}^\mathrm{cmb} \mathbf{\hat S}^\mathrm{cmb} \boldsymbol{\mathcal{A}}^{\mathrm{cmb},t})
    \Big\},
    \label{eq:S_no_priors}
\end{equation}
where:
\begin{equation}
\mathbf{F} \, \equiv \, \frac{1}{n_\mathrm{pix}}\,\boldsymbol{\mathcal{A}}^\mathrm{fg}\,\sum_p\, \mathbf{c}_\mathrm{p}^\mathrm{fg}\mathbf{c}_\mathrm{p}^{\mathrm{fg}, t} \boldsymbol{\mathcal{A}}^{\mathrm{fg},t}.
\end{equation}
The total ensemble average likelihood therefore takes into account average over noise realisations through the noise covariance matrix $\mathbf{N}$, and over CMB realisations through the single-pixel, CMB covariance matrix $\mathbf{\hat S}^\mathrm{cmb}$. In what follows, we use the expression Eq.~(\ref{eq:S_no_priors}) for the spectral likelihood, and therefore keep the assumption that $\boldsymbol{\mathcal{A}}$ and $\mathbf{N}$ are sky pixel independent.

\subsubsection{Priors on instrumental parameters}
\label{priors}

As mentioned earlier we expect that not all instrumental parameters can be constrained using the observed data with sufficient precision. Whenever this is the case, we may need to introduce calibration priors on some instrumental parameters either to break degeneracies in the system and/or to ensure a
tolerable level of residuals. Where appropriate, we add Gaussian priors to the ensemble average spectral likelihood as:
\begin{equation}
    \mathcal{S}' = \mathcal{S} + \sum_{\beta_h} \frac{1}{2 \sigma^2_{\beta_h}} \, (\beta_h - \Tilde{\beta_h})^2,
    \label{eq:S_priors}
\end{equation}
where $\mathcal{S}$ is the likelihood of Eq.~(\ref{eq:S_no_priors}) and $\sigma_{\beta_h}$ is the calibration error on parameter $\beta_h$. We assume the calibration measurement to be unbiased with a $1 \sigma$ error given by $\sigma_{\beta_h}$. Its best fit value will be typically different than the true value of the parameter, denoted with $\Tilde{\beta}_h$, but should be within the estimated uncertainty from it, $\sigma^2_{\beta_h}$. Our prior in Eq.~(\ref{eq:S_priors}) is assumed to be averaged over an ensemble of calibration procedures. Consequently, it is centered at the true value of the parameter and its uncertainty is larger by a factor of 2 (in quadrature) than that of a single calibration result.

\subsubsection{Statistical error}
\label{stat-error}

In this approach, the maximum-likelihood values obtained by minimizing Eq.~(\ref{eq:S_priors}) are average values, and we can compute the matrix of second derivatives at the peak of the likelihood --- the Hessian matrix $\mathcal{H}$:
\begin{equation}
    \mathcal{H}_{\beta\beta'} \equiv \left \langle \frac{\partial^2 \mathcal{S'}}{\partial \beta \, \partial \beta'} \Big |_\mathrm{peak} \right \rangle_{\mathrm{cmb}+\mathrm{noise}} 
     \label{eq:fisher},
\end{equation}
where we do not assume that the peak of the likelihood corresponds to true values of parameters $(\Tilde{\beta}_f,\Tilde{\beta}_h)$. The Hessian measures the curvature of the likelihood at its peak, and is directly related to the uncertainty due to instrumental noise and the CMB:
\begin{equation} 
\mathbf{\Sigma} \simeq \mathcal{H}^{-1}
\label{eq:sigma}
\end{equation}
This statistical error matrix $\mathbf{\Sigma}$ can be computed analytically following Eq.~(A5) in \cite{xforecast}.

If the assumed mixing matrix $\boldsymbol{\mathcal{A}}(\beta_f,\beta_h)$ corresponds to the true
$\boldsymbol{\mathcal{A}}$ for a given set of parameters $(\beta_f,\beta_h)$ derived from the minimisation of the spectral likelihood, the statistical uncertainty on the estimated parameters expressed by $\mathbf{\Sigma}$ will be the only source of residuals in the cleaned CMB map. Following~\cite{xforecast} we refer to those as statistical residuals. These increase the error on the estimated cosmological parameters but do not bias their estimated values \cite{errard2011}. However, if the assumed and true mixing matrices do not match, there will be systematic differences between the estimated and true sky components leading to systematic residuals in the cleaned CMB map and therefore potentially to biases in the estimated values of cosmological parameters~\cite{xforecast}.

In the next section, we derive the expression of both systematic and statistical residuals for the case of the generalised data model.

\subsection{Residuals}
\label{res}

Given a set of $(\beta_f,\beta_h)$ parameters, we can compute the corresponding mixing matrix $\boldsymbol{\mathcal{A}}$, and reconstruct the estimation of the noiseless component maps $\mathbf{\hat{c}}$ \cite{maxlike}:
\begin{equation}
    \mathbf{\hat{c}}_p = (\boldsymbol{\mathcal{A}}^t \, \mathbf{N}^{-1} \, \boldsymbol{\mathcal{A}})^{-1} \, \boldsymbol{\mathcal{A}}^t \mathbf{N} ^{-1} \, \mathbf{s}_p \equiv \mathbf{W}(\beta_f,\beta_h) \, \mathbf{s}_p,
\end{equation}
where we recall that $\mathbf{s}$ is the noiseless sky signal.
The noiseless residuals are defined as the difference between the reconstructed map and the true signal:
\begin{equation}
    \mathbf{r}_p \equiv \mathbf{\hat{c}}_p \, - \, \mathbf{c}_p = \mathbf{W}(\beta_f,\beta_h) \, \mathbf{s}_p \, - \, \mathbf{c}_p
    \label{eq:residuals-simple}
\end{equation}
We consider two main contributions to the residuals: (1) statistical residuals, reflecting the statistical scatter due the instrumental noise and CMB realizations; and (2) systematic residuals, reflecting the systematic (averaged) differences between the true sky and the best-fit of the assumed model. Rewriting Eq.~(\ref{eq:residuals-simple}) for the CMB channel only, we have:
\begin{eqnarray}
    \mathbf{r}^{\mathrm{CMB}}_p(\beta_f,\beta_h) & = &\underbrace{\mathbf{W}^0(\beta_f,\beta_h) \mathbf{F}_p}_{\mathrm{foregrounds \, residuals}} \nonumber\\ &+ &\;\; \underbrace{\mathbf{W}^0(\beta_f,\beta_h) \mathbf{C}_p \, - \,  \mathbf{c}_p^\mathrm{cmb}}_{\mathrm{CMB \, residuals}},
    \label{eq:cmb-res}
\end{eqnarray}
where the index $0$ denotes the CMB part of the $\mathbf{W}$ operator and $\mathbf{F}_p$ (resp.  $\mathbf{C}_p$) is the total foreground (resp. CMB) contribution at each frequency, gathering contribution from all mixed Stokes parameters ($\boldsymbol{\mathcal{C}}_0,\boldsymbol{\mathcal{S}}_0,\boldsymbol{\mathcal{C}}_4,\boldsymbol{\mathcal{S}}_4$) defined in Eq.~(\ref{eq:data-model-w-bp}).
Whenever elements of the mixing matrix corresponding to the CMB component are assumed to be known, as it was the case in the original application of this framework~\cite{xforecast,xForecast2}, the second term in Eq.~(\ref{eq:cmb-res}) vanishes and the recovered CMB map contains always all the CMB signal and is merely contaminated by the foreground residuals. However, this term has to be taken into account in the present case as the instrumental effects may change the way the CMB amplitude varies between the frequency bands. Here we therefore generalize the derivation of~\cite{xforecast} explicitly accounting for this term.

We can rewrite the CMB contribution of Eq.~(\ref{eq:cmb-res}) as:
\begin{align}
     \mathbf{W}^0 \mathbf{C} \, - \,  \mathbf{c}_\mathrm{cmb} & = \mathbf{W}^0 \,  \boldsymbol{\mathcal{A}}^\mathrm{cmb} \, \mathbf{c}_\mathrm{cmb} \, - \,  \mathbf{c}_\mathrm{cmb} \nonumber \\ 
     & = (\mathbf{W}^0 \, \boldsymbol{\mathcal{A}}^\mathrm{cmb}  \, - 1) \,  \mathbf{c}_\mathrm{cmb}.
\end{align}
This simplifies the estimation of CMB contribution in the residuals, that can be computed using the simplified procedure described in Appendix D of~\cite{xforecast}.

We perform a Taylor expansion of Eq.~(\ref{eq:cmb-res}) around the estimated best fit values for $\beta = \{\beta_h, \beta_f\}$, denoted $\hat{\beta}$ :
\begin{align}
    \mathbf{r}^{\mathrm{CMB}}_p(\beta) & \simeq \,  \mathbf{W}^0_p(\hat{\beta}) (\mathbf{F}_p + \mathbf{C}_p) \nonumber \\
    & \left. + \sum_\beta \delta \beta \frac{\partial \mathbf{W}^0_p}{\partial \beta} \right| _{\hat{\beta}}  (\mathbf{F}_p + \mathbf{C}_p)\nonumber \\
     & \left. + \sum_{\beta,\beta'} \delta \beta \, \delta \beta' \frac{\partial^2 \mathbf{W}^0_p}{\partial \beta \, \partial \beta'} \right| _{\hat{\beta}} (\mathbf{F}_p + \mathbf{C}_p) \nonumber \\
    &  \, - \mathbf{c}_p^\mathrm{cmb}, 
     \label{eq:res}
\end{align}
with $\delta \beta \equiv \beta - \tilde{\beta}$.

We introduce new quantities in pixel-domain as in~\cite{xforecast}:
\begin{eqnarray}
    \mathbf{y}_p & \equiv& \mathbf{W}^0_p(\hat{\beta}) (\mathbf{F}_p + \mathbf{C}_p) \, - \, \mathbf{c}_p^\mathrm{cmb}  \nonumber \\
     \mathbf{Y}^{(1)}_{p,\beta} &\equiv & \left. \sum_\beta \frac{\partial \mathbf{W}^0_p}{\partial \beta} \right| _{\hat{\beta}} (\mathbf{F}_p + \mathbf{C}_p) \nonumber \\
     \mathbf{Y}^{(2)}_{p,\beta \beta'} &\equiv & \left. \sum_{\beta,\beta'} \frac{\partial^2 \mathbf{W}^0_p}{\partial \beta \, \partial \beta'} \right| _{\hat{\beta}} (\mathbf{F}_p + \mathbf{C}_p),
\end{eqnarray}
so that we can rewrite Eq.~(\ref{eq:res}):
\begin{equation}
    \mathbf{r}^{\mathrm{CMB}}_p(\beta) \simeq \mathbf{y}_p \, + \, \sum_\beta \, \delta \beta \, \mathbf{Y}^{(1)}_{p,\beta}  \, + \, \sum_{\beta,\beta'} \, \delta \beta \, \delta \beta' \, \mathbf{Y}^{(2)}_{p,\beta\beta'}.
\end{equation}
We fit for a single value of every $\beta$ for all considered pixels and can rewrite the same equation in the harmonic domain and express the total level of foreground residuals as:
\begin{align}
    \mathcal{C}^\mathrm{res}_\ell &\equiv \boldsymbol{\otimes}_\ell(\mathbf{r}^{\mathrm{CMB}},\mathbf{r}^{\mathrm{CMB}}) \\
    &\simeq   \boldsymbol{\otimes}_\ell(\mathbf{y},\mathbf{y}) \, + \, \boldsymbol{\otimes}_\ell(\mathbf{y},\mathbf{z}) \, + \, \boldsymbol{\otimes}_\ell(\mathbf{z},\mathbf{y}) \nonumber \\
    & \, + \, \tr[\mathbf{\Sigma} \boldsymbol{\otimes}_\ell(\mathbf{Y}^{(1)}\, , \mathbf{Y}^{(1)})]
    \label{eq:cl_res}
\end{align}
where the symbol $\boldsymbol{\otimes}$ denotes the cross-spectrum of two quantities {\em averaged over statistical ensemble of the CMB realizations}~\footnote{We note that in the case studied in~\cite{xforecast} this average was trivial as there was no CMB contribution to the residuals and therefore usually omitted.}, and $\mathbf{z}$ is defined as:
\begin{equation}
    \mathbf{z}_p \equiv \tr [\mathbf{Y}_p^{(2)} \mathbf{\Sigma}].
    \label{eq:z}
\end{equation}
As outlined previously, residuals are composed of two main contributions, statistical and systematic residuals, as well as a cross-term:
\begin{align}
    \mathcal{C}^\mathrm{syst.}_\ell \equiv  & \boldsymbol{\otimes}_\ell(\mathbf{y},\mathbf{y})
    \label{eq:res_syst} \\
    \mathcal{C}^\mathrm{stat.}_\ell \equiv & \tr[\mathbf{\Sigma} \boldsymbol{\otimes}_\ell(\mathbf{Y}^{(1)}\, , \mathbf{Y}^{(1)})].
     \label{eq:res_stat} \\
     \mathcal{C}^\mathrm{cross}_\ell \equiv & \boldsymbol{\otimes}_\ell(\mathbf{y},\mathbf{z}) \, + \, \boldsymbol{\otimes}_\ell(\mathbf{z},\mathbf{y}).
     \label{eq:res_cross}
\end{align}
We can also write the full covariance matrix of the CMB map, $\mathbf{\tilde{c}}^\mathrm{cmb} (\equiv \mathbf{c} \, + \, \mathbf{r}^\mathrm{CMB} \, + \, \mathbf{n})$, as recovered via the component separation procedure described here. It reads,
\begin{eqnarray}
\mathbf{E} & \equiv & \langle \mathbf{\tilde{c}}^\mathrm{cmb}\,\mathbf{\tilde{c}}^\mathrm{cmb}\rangle_{\mathrm{cmb}+\mathrm{noise}} \, 
 = \, \mathbf{C}^\mathrm{cmb} \, + \, \mathbf{N} \, + \, \langle \mathbf{y}\mathbf{y}^\dagger\rangle_{\mathrm{cmb}} \nonumber \\
& + &
\langle \mathbf{z}\mathbf{y}^\dagger\rangle_{\mathrm{cmb}} \, +
\, \langle \mathbf{y}\mathbf{z}^\dagger\rangle_{\mathrm{cmb}} \, + \,
\, \langle \mathbf{Y}^{\left(1\right)}\mathbf{\Sigma}\mathbf{Y}^{\left(1\right),\;\dagger}\rangle_{\mathrm{cmb}}, \ \ \ \ \ 
\label{eq:Edef}
\end{eqnarray}
where $\mathbf{C}^\mathrm{cmb}$ denotes the covariance of the CMB signal, $\mathbf{N}$ that of the noise in the recovered map, and the angle brackets denote averaging over statistical ensemble as defined by the subscript.
\subsection{Noise}
\label{noise}

We assume that the noise is homogeneous and uncorrelated for all frequency channels, and given that we have one value of all parameters $\beta$ for the entire sky patch we can express the noise power spectrum in the cleaned CMB map as in Eq.~(32) of \cite{xforecast}:

\begin{equation}
    \mathcal{C}^\mathrm{noise}_\ell = [(\boldsymbol{\mathcal{A}}^t \, \mathbf{N}^{-1}_\ell \, \boldsymbol{\mathcal{A}})^{-1}]_{\mathrm{CMB \, \times \, CMB}}
\end{equation}
with $\mathbf{N}_\ell$ describing the noise spectra of each frequency map, taking its resolution into account:

\begin{equation}
    \mathbf{N}^{ij}_\ell \equiv (w_i)^{-1} \exp \Big( \ell \, (\ell + 1) \, \frac{\mathrm{FWHM}^2_i}{8 \, \log 2} \Big) \delta^j_i
    \label{eq:noise}
\end{equation}
with $(w_i)^{-1/2}$ the sensitivity, and $\mathrm{FWHM}_i$ the full-width half maximum of the corresponding frequency band $i$.

\subsection{Cosmological likelihood}
\label{cosmo-likelihood}

Following Appendix C2 in \cite{xforecast}, we write the cosmological parameter likelihood, averaged over instrumental noise and CMB signal realisations:
\begin{equation}
    \langle S^\mathrm{cos} \rangle \equiv \tr \mathbf{C}^{-1} \mathbf{E} \, + \, \ln \det \mathbf{C}
    \label{eq:cosmo-likelihood}
\end{equation}
where $\mathbf{C}$ is the assumed multi-pixel covariance of the cleaned CMB signal, and $\mathbf{E}$ is the multi-pixel correlation matrix of the CMB map retrieved with the component separation procedure as defined in Eq.~(\ref{eq:Edef}). Consequently, $\mathbf{E}$ takes into account the presence of the residuals, Eqs.~(\ref{eq:res_syst}), (\ref{eq:res_stat}), and ~(\ref{eq:res_cross}), and all which can be computed semi-analytically given the actual model of the data.

The assumed covariance $\mathbf{C}$ expresses the state of our knowledge about the data. Ideally, we would wish that $\mathbf{C} = \mathbf{E}$ however in practice this is rarely the case. In the following we will consider two cases. In the first case, the assumed covariance is that of the CMB signal only, thus ignoring entirely the effects of the component separation. We have therefore:
\begin{equation}
\mathbf{C} = \mathbf{C}^\mathrm{cmb} \, + \, \mathbf{N}.
\label{eq:no_deprojection}
\end{equation}
In the second case we assume that the statistical errors can be modelled on some level. Specifically, we assume that:
\begin{equation}
\mathbf{C} = \mathbf{C}^\mathrm{cmb} + \mathbf{C}^\mathrm{stat.} \, + \, \mathbf{N},
\label{eq:deprojectCov}
\end{equation}
where $\mathbf{C}^\mathrm{stat.}$ is the covariance matrix of the statistical residuals $C_\ell^\mathrm{stat.}$ defined in Eq.~(\ref{eq:res_stat}). We note that sufficient information allowing for effective modelling of the statistical residuals may be indeed available either internally, in some self-consistent statistical approaches, e.g.,~\cite{spectral-likelihood,commander}, or using some external data, e.g.,~\cite{planck_likelihood}. Hereafter, we refer to this second case as the deprojection case. Detailed expressions are given in Appendix \ref{deprojection}.

\section{Application}
\label{results}

In this section, we first introduce the instrumental configuration and parametrisation that we adopt in this work (\ref{inst-config}), then we present various sets of parameters that we consider and discuss uncertainties and degeneracies on parameters in these various cases (\ref{cases}). From these, we can finally estimate foreground residuals and forecast their impact on tensor-to-scalar ratio determination (\ref{fg_res}).

\subsection{Configuration}
\label{inst-config}

The framework presented here is very flexible and could be adapted to a broad range of experiment designs with different instrument models. For definiteness, we demonstrate it in the case of a typical, ground-based, Stage 3 CMB polarisation experiment. The configuration and parameters we use are motivated by the publicly available Simons Observatory Small Aperture Telescopes design \cite{SO}. However, this is not the goal of this work to make reliable performance forecasts for any particular experiment. Here, we use this setup merely for demonstrative purposes.

\subsubsection{Telescopes and frequency bands}

As outlined in introduction, to enable for component separation, modern CMB polarisation experiments typically deploy several instruments to trace foregrounds at different frequencies, so that we can disentangle them from the black body CMB signal. In several ground-based experiments, these frequency bands are grouped by two on the same focal plane, hence the use of an achromatic HWP that has to accommodate two different observing frequencies in the same optics tube. Following the Simons Observatory configuration \cite{SO}, we consider three telescopes, whose frequency coverage is as follows:
\begin{itemize}
    \item Low Frequency (LF) telescope: frequency bands centered at 27 GHz and 39 GHz;
    \item Mid Frequency (MF) telescope: frequency bands centered at 93 GHz and 145 GHz;
    \item Ultra-High Frequency (UHF) telescope: frequency bands centered at 225 GHz and 280 GHz.
\end{itemize}
We model input noise covariance matrices in the pixel domain as
\begin{equation}
     \mathbf{N}^{ii}_p = w_i^{-1},
\end{equation}
with $w_i^{-1}$ the sensitivity of the corresponding frequency band $i$. We use the publicly available sensitivity calculator~\footnote{\texttt{SO\_Noise\_Calculator\_Public.py} available at \url{ https://simonsobservatory.org/publications.php}} to compute realistic instrumental sensitivities corresponding to the frequency coverage.

\subsubsection{HWP}
\label{hwp-config}
For each telescope (LF, MF, UHF), we consider a typical 3-layer achromatic HWP, and that only the central layer is rotated with respect to the reference frame of the instrument, i.e. $\alpha_1 = \alpha_3 = 0$, as for example in \cite{PB-HWP}. This typical choice of angle is driven by the maximisation of transmitted power and is known as Pancharatnam design \cite{achromatic-hwp}. This could be straightforwardly generalised to other cases, for example for an HWP with more than 3 layers, as proposed for the LiteBIRD mission which will deploy a 9 layers on its low-frequency telescope~\cite{LB-HWP}.

The Mueller matrix of the 3-layer HWP that we consider can therefore be written as:
\begin{equation}
   \mathbf{M}_\mathrm{HWP}=\mathbf{M}_\mathrm{layer}\mathbf{R}(-2\alpha_2)\mathbf{M}_\mathrm{layer}\mathbf{R}(2\alpha_2)\mathbf{M}_\mathrm{layer},
    \label{eq:M_HWP}
\end{equation}
with $\mathbf{M}_\mathrm{layer}$ defined as in Eq.~(\ref{eq:one-layer}). The Mueller matrix of a single layer is parametrised by $\delta$, Eq.~(\ref{eq:delta}). As already pointed out in section \ref{hwp}, once the birefringent material is chosen, $\delta$ depends on observing frequency, $\nu$, and thickness of each layer, $\theta_\mathrm{hwp}$. For a given HWP, we assume all layers to be identical, and parametrize $\mathbf{M}_\mathrm{layer}$ with only its thickness $\theta_\mathrm{hwp}$.

Based on existing optimised parameters \cite{tomo} and proposed designs \cite{SO,HWP-SPIE,PB-HWP}, we choose the following nominal values for HWP parameters:
\begin{itemize}
    \item $\theta_\mathrm{hwp}(\mathrm{LF}) =$ 14.36 mm
    \item $\theta_\mathrm{hwp}(\mathrm{MF}) =$ 3.8 mm    \item $\theta_\mathrm{hwp}(\mathrm{UHF}) =$ 1.86 mm
    \item $\alpha_2$ = 58\textdegree
\end{itemize}
We note here that $\theta_\mathrm{hwp}$ is different for each telescope since the thickness is adapted to the frequency range of observation, when $\alpha_2$ is in principle the same. However, to account for possible fabrication and/or calibration differences between HWP, we allow $\alpha_2$ to vary for each of the telescopes when fitting for parameters (although they have the same nominal value).

Details of $\boldsymbol{\mu}_{ij}$, the coefficients of the HWP Mueller matrix, $\mathbf{M_{\rm HWP}}$, as functions of $\alpha_2$ and $\theta_\mathrm{hwp}$ can be found in Appendix \ref{3layer-appendix}.

\subsubsection{Bandpasses}
\label{bp-design}

We parametrize each bandpass by two parameters: its center $\nu_0$ and its half-width $\Delta \nu$. For the mock data we assume  the standard $30$\% bandpass bandwidths for all frequency bands and therefore set $\Delta \nu = 0.15 \times \nu_0$. In the model, we allow $\nu_0$ and $\Delta \nu$ to vary independently.

\begin{figure}
    \centering
    \includegraphics[width=0.5\textwidth]{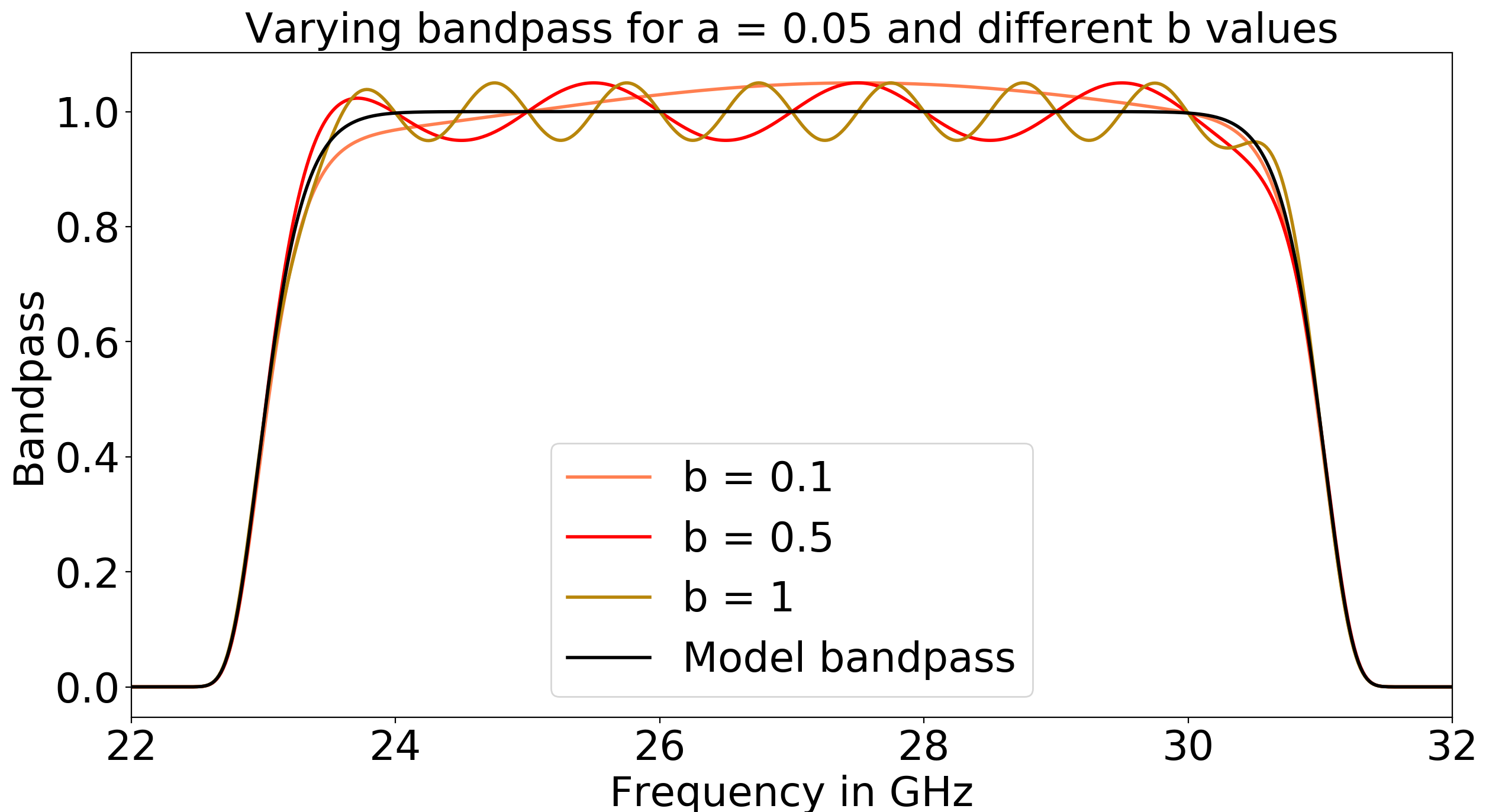}
    \caption{Ideal and varying bandpass model for the $27$GHz channel, for $a= 0.05$ and different $b$ values following Eq.~(\ref{eq:bp-mod}). We first use the ideal bandpass in both data and model, and we then introduce a varying bandpass in data only, to study the impact of this mismatch on systematic residuals, as detailed in section \ref{r-bp}}
    \label{fig:bp}
\end{figure}

We model the bandpass as top-hat function with smoothed edges as shown in Figure \ref{fig:bp}. 
The analytic expression of the bandpass $\mathcal{B}(\nu)$ as a function of $\nu_0$ and $\Delta\nu$ is
\begin{equation}
   \mathcal{B}(\nu) = \exp{\left[- \left( \frac{ |\nu-\nu_0|}{\Delta \nu} \right) ^{20} \right]}.
   \label{eq:bp}
\end{equation}
Realistic parametric modeling of bandpass can be complex because true bandpasses are never as regular as the one shown in Figure \ref{fig:bp}, see e.g.~\cite{ward2018}. The idealized model in Eq.~(\ref{eq:bp}) is however convenient as it permits studying the impact on the results of the two main bandpasses characteristics: the band center and its width. 

Concurrently, we also explore the effects of potential deviations from this simplified bandpass form. This can reflect more complex intrinsic
shape of the realistic bandpasses but also be due to detector-to-detector variations. To this end we propose a toy-model given by:
\begin{equation}
    \mathcal{B}'(\nu) =  \mathcal{B}(\nu) \times \left[ 1 - a \times \sin{(b \times 2 \pi \nu)} \right],
    \label{eq:bp-mod}
\end{equation}
and which includes deviations from the idealized, ``average'', form in Eq.~(\ref{eq:bp}). An example of such bandpasses is shown in Figure \ref{fig:bp}.

In the following, we first use the simplified bandpass model, $\mathcal{B}(\nu)$, in both the input data and the model, to assess the impact of uncertainty in the bandpass position and width on component separation. 
Subsequently, we use the more complex form, $\mathcal{B}'(\nu)$, to generate input data, but we continue fitting the simple model of Eq.~(\ref{eq:bp}) to the data, thus neglecting the presence of the small-scale features in the bandpasses, which are left unmodeled. As pointed out in section \ref{res}, a mismatch between the data and the fitted model will typically result in systematic residuals in reconstructed maps. As mentioned earlier such a mismatch could arise due to detector-to-detector variability of the bandpasses. We therefore explore the impact on systematic residuals level and cosmological parameters of various $a$ and $b$ values in section \ref{r-bp}, trying to understand what level of unmodeled deviations, in terms of their scale or their amplitude, can be tolerated and at what point they need to be modeled (or measured in advance).

\subsubsection{Summary of instrumental parameters}

Table \ref{tab:params} summarises the hardware parameters that we consider in this model and their nominal values, used to generate the input data. We recall here that all parameters are independent and can in principle all vary when we minimize the likelihood of Eq.~(\ref{eq:S_priors}). 

\begin{table}
    \centering
    \begin{tabular}{|c|c|c|c|c|c|c|c|}
    \hline
        \multicolumn{2}{|c|}{  }  & \multicolumn{2}{c|}{LF}  & \multicolumn{2}{c|}{MF}  & \multicolumn{2}{c|}{UHF} \\
   \cline{3-8} 
    \multicolumn{2}{|c|}{  }  & LF1  & LF2  & MF1  & MF2  & UHF1  & UHF2  \\
    \hline
    \multirow{2}*{HWP} & $\alpha_2$ & \multicolumn{2}{c|}{58\textdegree} & \multicolumn{2}{c|}{58\textdegree}& \multicolumn{2}{c|}{58\textdegree} \\
    \cline{2-8} 
    & $\theta_\mathrm{hwp}$ & \multicolumn{2}{c|}{14.36 mm} & \multicolumn{2}{c|}{3.8 mm} & \multicolumn{2}{c|}{1.86 mm} \\
    \hline
    
    \multirow{2}*{Bandpass (GHz)} & $\nu_0$ & 27 & 39 & 93 & 145 & 225 & 280 \\
    \cline{2-8} 
    & $\Delta \nu$ & 4.05 & 5.85 & 13.95 & 21.75 & 33.75 & 42.0 \\
    \hline
    \end{tabular}
    \caption{Nominal values of instrumental parameters}
    \label{tab:params}
\end{table}

In contrast, while we include in our data models parameters describing sinuous antennas, section~\ref{sin-antenna}, those are fixed throughout this analysis.   
Similarly, although beam modeling and beam systematics could also be investigated within this framework, we leave this for future work and do not assume any beam smoothing of input maps. Note that beam smoothing is taken into account in the harmonic domain when we estimate noise after component separation in Eq.~(\ref{eq:noise}). Again, beam sizes that we adopt here correspond roughly to the Simons Observatory Small Aperture Telescopes ones \cite{SO}.

\subsubsection{Input sky}

\label{sec:mapParams}

We use the \texttt{PySM} package~\cite{pysm} to generate foregrounds, and the \texttt{healpy} package~\cite{healpy} to generate CMB maps from fiducial CMB power spectra. We choose the following reference frequency for the foregrounds templates:

\begin{align}
    \nu_{0,\mathrm{sync.}} & = 70 \, \mathrm{GHz}, \ \ \ \ \ \ 
    \nu_{0,\mathrm{dust}} = 353 \, \mathrm{GHz}
\end{align}
We fix the dust temperature at $T = 19.6$~K and for spectral indices we consider the nominal values:

\begin{align}
    \beta_d & = 1.59, \ \ \ \ \ \ 
    \beta_s = -3.1.
\end{align}
We take constant foreground parameters across the sky, although the effect of varying spectral parameters could be included within our framework. For details we refer the reader to~\cite{xForecast2}.

\texttt{PySM} uses the \texttt{healpy} implementation of the \textsc{HEALP}ix pixelisation scheme~\footnote{\url{https://healpix.sourceforge.io/}}, and we use \texttt{nside} = 256 throughout this work. We use a mask corresponding to 10\% of the sky. When it comes to power spectrum reconstruction, we therefore consider multipoles from $\ell = 30$ to $\ell \sim 500$, given the limitation by the mask at low $\ell$, and by the pixel size at high $\ell$.

We consider two cosmological parameters, the tensor-to-scalar ration $r$ which sets the amplitude of primordial B modes:

\begin{equation}
    C_{\ell,\,\mathrm{primordial}}^{BB} = r \times C_{\ell,\,\mathrm{primordial}}^{BB}(r = 1),
\end{equation}
and the lensing parameter which sets the delensing amplitude ($A_L = 1$ corresponds to no delensing and $A_L = 0$ corresponds to full delensing): 

\begin{equation}
        C_{\ell,\,\mathrm{lensing}}^{BB} = A_L \times C_{\ell,\,\mathrm{lensing}}^{BB}(A_L = 1).
\end{equation}
We choose
\begin{align}
    r & = 0, \ \ \ \  \ \ 
    A_L = 1,
\end{align}
as our reference case but also explore other cases in section~\ref{r}.

Finally, as we aim at modeling a ground-based experiment, we consider that sky-only modulated terms, i.e. $\mathcal{C}_0$ and $\mathcal{S}_0$ in our model, will be compromised by the atmospheric noise at long temporal modes of the collected data streams~\cite{abs-hwp,PB-HWP-2}. In what follows, we therefore use only $\mathcal{C}_4$ and $\mathcal{S}_4$, in the simulated data and in the model. However, we point out that, for a space mission such as LiteBIRD, it could be possible to recover all four mixed Stokes components.

\subsection{Optimization of the generalized spectral likelihood}
\label{cases}

In this section, we explore the parameter space described in the previous section: 18 instrumental parameters (6 for the HWPs and 12 for the bandpasses), as well as 2 foreground spectral parameters $\beta_s$ and $\beta_d$. We start by fixing all instrumental parameters to their nominal values, and we only estimate foreground parameters, $\beta_d$ and $\beta_s$. This constitutes our reference scenario for uncertainties, residuals and bias on $r$, and we compare results obtained when estimating instrumental parameters to this case. We then progressively free instrumental parameters in the spectral likelihood, Eq.~(\ref{eq:S_no_priors}). We do not consider priors in the first instance, but will introduce them for specific cases whenever it becomes necessary.

We consider the following cases, and corresponding abbreviations to identify them throughout this work:

\begin{itemize}
    \item Spectral Energy Distribution (SED) parameters (foreground spectral indices) only 
    \\ $\rightarrow$ SED only
    \item SED + HWP central layer angle for all three HWP \\ $\rightarrow$ SED + $\alpha_2$
    \item SED + HWP layer thickness for all three HWP 
    \\ $\rightarrow$ SED + $\theta_\mathrm{hwp}$
    \item SED + HWP central layer angle + HWP layer thickness for all three HWP
    \\ $\rightarrow$ SED + HWP
    \item SED + Bandcenters for all bandpasses
    \\ $\rightarrow$ SED + $\nu_0$
    \item SED + Bandwidths for all bandpasses
    \\ $\rightarrow$ SED + $\Delta \nu$
    \item SED + Bandcenters + Bandwidths for all bandpasses 
    \\ $\rightarrow$ SED + Bandpass
    \item SED + all the above 
    \\ $\rightarrow$ SED + All
\end{itemize}

\subsubsection{Method}
For each case, once the generalized spectral likelihood is optimized, we compute the Hessian matrix $\mathcal{H}$ as defined in Eq.~(\ref{eq:fisher}) at the numerically-determined peak of the likelihood. We compare the one-dimensional spectral likelihood where we fix all parameters but one to their nominal value, to a Gaussian function whose variance is determined by the diagonal of the Hessian matrix:

\begin{equation}
    \sigma_i = \frac{1}{\sqrt{\mathcal{H}_{ii}}}
\end{equation}
for any parameter $i$.

We also compute the eigenvalues and eigenvectors decomposition of the Hessian matrix, to test for the presence of possible degeneracies in the considered parameter space.

Finally, we evaluate the marginalised error bar on spectral parameter using the diagonal of the inverse of the Hessian matrix, $\mathbf{\Sigma}$, as defined in Eq.~(\ref{eq:sigma}):

\begin{equation}
    \sigma(\beta_i) = \sqrt{\mathbf{\Sigma}_{\beta_i,\beta_i}}
    \label{eq:marg-sigma}
\end{equation}
The errors on instrumental and foreground parameters are of primary importance as they determine the amplitudes of both statistical and systematic residuals which impact directly our estimates of $r$. Generically, we expect that uncertainty on these parameters will increase as we increase the number of free instrumental parameters. 

We emphasise however that there are fundamental differences in the way the data constrain both these types of parameters, and therefore the number of parameters that can be constrained.
Indeed, the foreground parameters are usually specific to a sky component, be that dust or synchrotron, and therefore tend to affect only a single column of the generalized mixing matrix. In contrast, the instrumental parameters are rather frequency channel specific and determine the rows of the matrix. As a consequence, the constraints on the instrumental parameters do not benefit from the multi-frequency information as much as do the foregrounds parameters. In particular, the quality of the constraints do not improve with the increasing number of frequency bands (while the number of parameters will typically increase). In the simple case of instrumental parameters specific to only a single frequency channel we expect that only one such parameter can be well constrained with the data. However, as we measure a few pieces of information per frequency channel, corresponding to the mixed-Stokes maps which we consider for each frequency channel, we could in principle set more than a single constraint, and up to as many as the available pieces of information. Most of these constraints, in fact all but one, will typically be of significantly lower precision and often insufficient for our purpose, rendering only one, useful constraints on a linear combination of the considered parameters. If instrumental parameters are relevant to more than a single frequency channel the number of potentially well constrained parameters increases. We note that these analytic insights are fully borne-out by our numerical results as we discuss in the following section.

\subsubsection{HWP parameters}
We first consider the following cases, involving only HWP parameters:
\begin{itemize}
    \item SED + $\alpha_2$
    \item SED + $\theta_\mathrm{hwp}$
    \item SED + HWP
\end{itemize}

We find that all the likelihoods corresponding to these cases are very well approximated by a Gaussian, with the dispersion obtained by the corresponding inverse Hessian calculated at the peak of each likelihood. Moreover, the Hessians are very well conditioned, and all the parameters in all these cases are consequently well constrained and only weakly correlated. 

These observations apply to conditional and marginalized likelihood. We then consequently use the Gaussian approximation and the Hessians to compute the marginalised uncertainty on spectral parameters (defined in Eq.~(\ref{eq:marg-sigma})) in these various cases. We find that there is no significant increase of the marginalised uncertainty on spectral indices when adding HWP parameters, as shown in Table \ref{tab:sigma_beta_all} and Figure \ref{fig:sigma_beta}.

We thus expect that when only HWP parameters are considered, the impact on foreground residuals and $r$ estimation should be limited, as we will demonstrate in section \ref{fg_res}. We therefore conclude that in none of these cases we need priors to properly constrain HWP parameters. Though this conclusion is drawn in the case of the specific model of the HWP and the specific experimental set-up, we expect that it will hold as long as the number of HWP parameters does not exceed the number of frequency channels in the corresponding focal plane, otherwise some prior knowledge may become necessary.

\subsubsection{Bandpass parameters}
\label{bp-params} 
In the case of bandpasses we consider the following cases:

\begin{itemize}
    \item SED + $\nu_0$
    \item SED + $\Delta \nu$
    \item SED + Bandpass
\end{itemize}

\begin{figure*}
    \centering
    \includegraphics[width=\textwidth]{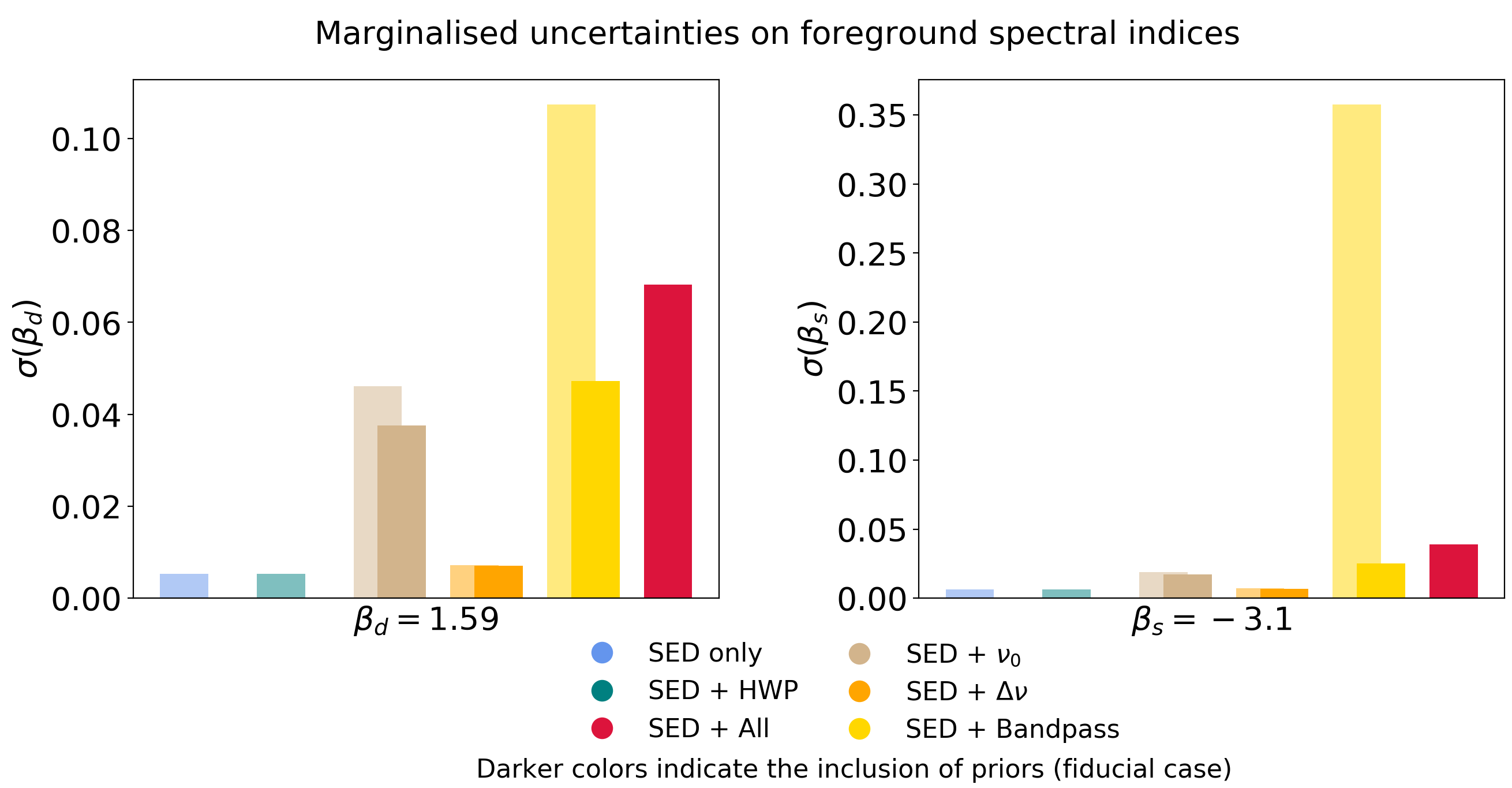}
    \caption{Marginalised 1-$\sigma$ uncertainties on foreground spectral indices. The inclusion of HWP parameters has a negligible impact on the uncertainty on spectral parameters, which is why we do not consider the inclusion of priors. However, the effect is much more sensitive when including bandpass parameters, thus the inclusion of priors. The uncertainty on spectral parameters when considering only $\nu_0$ or $\Delta \nu$ does not change much when adding priors, as shown in Table \ref{tab:sigma_beta_all}. However, we show that adding priors when all bandpass parameters are considered is required to limit uncertainty on foreground parameters and avoid degeneracies for instrumental parameters. In the case when all parameters are considered, priors are necessary because the system is otherwise degenerate.}
    \label{fig:sigma_beta}
\end{figure*}

\paragraph{Without priors}
In the cases with only either bandcenters or bandwidths in addition to foreground parameters allowed to change, i.e \{SED + $\nu_0$\} and \{SED + $\Delta \nu$\}, the system is not degenerate: its Hessian matrix is still formally positive definite as all eigenvalues are strictly positive. The spectral likelihood is well approximated by a Gaussian in these cases and we use the Hessian to set the constraints. We note that though the constraints on the bandpass centers are more precise relative to the parameter values than those obtained for the bandwidths, the impact of the former on the spectral parameters errors is significantly more pronounced. This is shown in Figure \ref{fig:sigma_beta}. We discuss the impact of this observation on the shape of foreground residuals in the next section.

When all bandpass parameters are allowed to vary at the same time, i.e. case \{SED + Bandpass\}, the system becomes nearly degenerate as we have two independent parameters for each frequency channel. In this case, the uncertainty on both spectral and instrumental parameters increases significantly. The Hessian matrix is still positive definite and errors on all parameters can be formally set, however, the eigenvalues of the Hessian matrix span now a range from $~10^4$ to $~10^{-2}$, rendering some of the constraints weak. As expected, the marginalised errors on spectral parameters largely increase in this case, as shown in Figure~\ref{fig:sigma_beta} and Table~\ref{tab:sigma_beta_all}.

\begin{figure}
    \centering
    \includegraphics[width=0.5\textwidth]{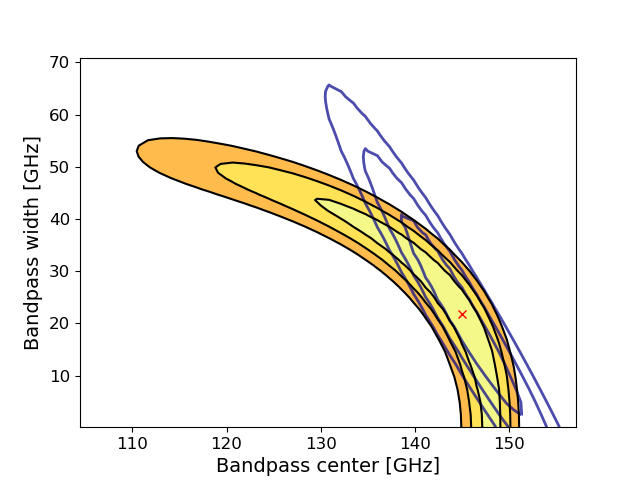}
    \caption{Two-dimensional gridding of the generalised spectral likelihood, Eq.~(\ref{eq:S_no_priors}), estimated from 100 random sky pixels. We overplot in blue the Gaussian approximation given by the Hessian, and show that in this case it fails to describe the system. The red cross indicate true values of parameters.}
    \label{fig:banana}
\end{figure}

 The near degeneracies in this case appear between a bandcenter and bandwidth of the same frequency band. This is illustrated by the elongated shape of the likelihood in a 2d space $(\nu_0, \Delta \nu)$ for the same bandpass as shown in Figure \ref{fig:banana}. We note that the Gaussian approximation fails in these cases to reproduce the behavior of the actual likelihood, but only in the ill-constrained directions. For the well-constrained directions, the Gaussian approximation and the Hessian continue however providing a very good description.

We thus conclude that adding bandpass parameters significantly increases uncertainties on both spectral and instrumental parameters, in particular in comparison with the case of the HWP parameters, impact of which was found to be very minor. We therefore consider the inclusion of calibration priors on bandpass parameters to alleviate this issue.

\paragraph{With priors}
\label{with_priors}

To remove bandpass-associated degeneracies in the parameter space, we introduce Gaussian priors on $\nu_0$ and $\Delta \nu$ as proposed in Eq.~(\ref{eq:S_priors}).

In general, the eigenvalues decomposition of the Hessian matrix provides good guidance on how to introduce priors in a most efficient and economic way. We start from the lowest eigenvalues and candidate parameters are those which are most aligned with the corresponding eigenvector. In the specific case here, the lowest eigenvalues of the Hessian matrix are associated with eigenvectors with dominant components corresponding to the bandpass parameters of the same frequency band. This is illustrated in Figure \ref{fig:banana}. As we found that the eigenvectors components in the directions of the band center and widths are comparable, we decided to set prior constraints on both of them for each band. We note that while other priors can also be considered here and may be helpful, the bandpass parameter degeneracies are the only degeneracies found in the studied problem and therefore the priors on these parameters are the most efficient and have the most pronounced impact.

In the absence of calibration data, we test three levels of the priors, in order to best illustrate the impact of them on the conclusions. These are:
\begin{itemize}
    \item Pessimistic: 5\% on $\nu_0$ and 8\% on $\Delta \nu$;
    \item Fiducial: 1\% on $\nu_0$ and 5\% on $\Delta \nu$;
    \item Optimistic: 0.5\% on $\nu_0$ and 1\% on $\Delta \nu$.
\end{itemize}

The corresponding uncertainties on spectral parameters for these different choices are listed on Table \ref{tab:bp_priors}.

\begin{table}
    \centering
    \begin{tabular}{|c|c|c|c|}
    \hline
         & Priors & $\sigma(\beta_d = 1.59)$ & $\sigma(\beta_s = -3.1)$ \\
    \hline
        SED only & No & $5.291 \times 10^{-3}$ & $6.420 \times 10^{-3}$ \\
     \hline
        SED + $\alpha_2$ & No & $5.292 \times 10^{-3}$ & $6.423 \times 10^{-3}$ \\
    \hline
        SED + $\theta_\mathrm{hwp}$ & No & $5.292 \times 10^{-3}$ & $6.422 \times 10^{-3}$ \\
    \hline
        SED + HWP & No & $5.300 \times 10^{-3}$ & $6.444 \times 10^{-3}$\\
    \hline
        SED + $\nu_0$ & No & $4.609 \times 10^{-2}$ & $1.890 \times 10^{-2}$ \\
    \hline
        SED + $\nu_0$ & \textbf{Yes} & $ 3.761 \times 10^{-2}$ & $1.736 \times 10^{-2}$ \\
    \hline
        SED + $\Delta \nu$ & No & $7.155\times 10^{-3}$ & $7.129 \times 10^{-3}$ \\
    \hline
        SED + $\Delta \nu$ & \textbf{Yes} & $7.014 \times 10^{-3}$ & $6.538 \times 10^{-3}$ \\
    \hline
        SED + Bandpass & No & $1.074 \times 10^{-1}$ & $3.577 \times 10^{-1}$\\
    \hline
     SED + Bandpass & \textbf{Yes} & $4.721 \times 10^{-2}$ & $2.493 \times 10^{-2}$\\
     \hline
       SED + All & No & Degenerate & Degenerate \\
    \hline
        SED + All & \textbf{Yes} & $6.825 \times 10^{-2}$ & $3.903 \times 10^{-2}$ \\
    \hline
    \end{tabular}
    \caption{Summary of 1-$\sigma$ marginalised uncertainties on spectral parameters. Priors refer to the fiducial case: 1\% on bandcenters ($\nu_0)$ and 5\% on bandwidths ($\Delta \nu$).}
    \label{tab:sigma_beta_all}
\end{table}

\begin{table}
    \centering
    \begin{tabular}{|c|c|c|c|}
    \hline
     & Priors & $\sigma(\beta_d = 1.59)$ & $\sigma(\beta_s = -3.1)$ \\
    \hline
     \multirow{4}*{SED + Bandpass} & No   &  0.108 & 0.358\\
    \cline{2-4}
     & Pessimistic & 0.0690  & 0.0419 \\
      \cline{2-4}
     & Fiducial & 0.0472 & 0.0249 \\
      \cline{2-4}
     & Optimistic & 0.0281 & 0.0161 \\
     \hline
    \end{tabular}
    \caption{Marginalised 1-$\sigma$ uncertainties on spectral parameters when considering bandpass parameters for different choices of priors.}
    \label{tab:bp_priors}
\end{table}

We also show uncertainties with and without priors on Figure \ref{fig:sigma_beta} for the fiducial case. The priors are efficient to reduce uncertainties in all considered cases, but their effect is most pronounced whenever all bandpass parameters are allowed to vary. These are indeed the cases when the system is nearly degenerate --- and the Hessian nearly singular. Introducing the priors reduces the range of eigenvalues by at least one or two orders of magnitude as does the condition number of the Hessian matrix.

When it comes to the level of priors, even not very accurate priors allow for a significant reduction of marginalised uncertainties on spectral parameters. However, the change in eigenvalues is not significant between pessimistic and fiducial cases, although it is better than in the no prior case. In what follows, unless otherwise is specified, we choose to keep the fiducial values as our nominal level of the priors, as they are closer to currently achieved calibration performance.

This choice of priors is not unrealistic at least in the cases of limited number of detectors, for instance compared to calibration performance currently achieved with a Fourier Transform Spectrometer (FTS) on a typical CMB experiment. Accuracy of such measurements has been demonstrated to be at the level of 1 GHz for the POLARBEAR experiment \cite{pb-fts} for example. Future experiments such as LiteBIRD plan on even higher accuracy on bandpass calibration, up to 0.2 GHz \cite{LB-bp}. The priors at the level we assume here should therefore be available for future analyses.

Overall, this shows that the inclusion of the priors at the level we propose is necessary and efficient to reduce marginalised uncertainties and lift degeneracies in the multi-parameter space. We also show in the next section that it reduces foreground residuals and bias on $r$.

\subsubsection{All parameters}
We consider here the case of,
\begin{itemize}
 \item SED + All
\end{itemize}
when all instrumental and foreground parameters are allowed to vary.
In this case, if we do not include priors, the Hessian matrix is nearly singular with some of the eigenvalues numerically zero.
This is expected given the large number of parameters. We therefore once again include the priors on bandpass parameters as discussed in the previous section, which resolve the degeneracy as shown in Table~\ref{tab:sigma_beta_all}. In the follow-up, unless otherwise specified, in case \{SED + All\} we always consider the fiducial priors on the bandpass parameters.

\subsubsection{Summary}

In this first part of our work, we have studied parameter space degeneracies as well as statistical uncertainties on spectral parameters estimated by the Hessian matrix at the peak of the spectral likelihood. We showed that we can estimate HWP parameters from the actual data with no significant increase of marginalised uncertainties, and without adding priors. This is due to the fact that HWP parameters apply to more than one frequency band, and therefore we have enough leverage to constrain them. However, when it comes to bandpass parameters, we showed that it is necessary to introduce priors on both bandcenters and bandwidths to avoid degeneracies in the parameter space and maintaining marginalised error bars on spectral parameters low enough. This is due to the fact that we have two bandpass parameters (bandwidth and bandcenter) for only one frequency band, and therefore these two parameters are degenerate. In what follows, unless otherwise is specified, we keep the following priors on bandpass parameters anytime these parameters are estimated: 1\% on $\nu_0$ and 5\% on $\Delta \nu$.

Overall, and as it could have been expected, we show that the more parameter we consider in the spectral likelihood, the higher statistical uncertainties, however different instrumental parameters affect the uncertainties on spectral parameters to a different degree. All results of this Section are summarised in Table \ref{tab:sigma_beta_all}. All these factors determine the level and shape of foreground residuals and the estimation of $r$, as we will now demonstrate. 

\subsection{Residuals}
\label{fg_res}

\subsubsection{Statistical residuals}

\begin{figure*}
    \centering
    \includegraphics[width=\textwidth]{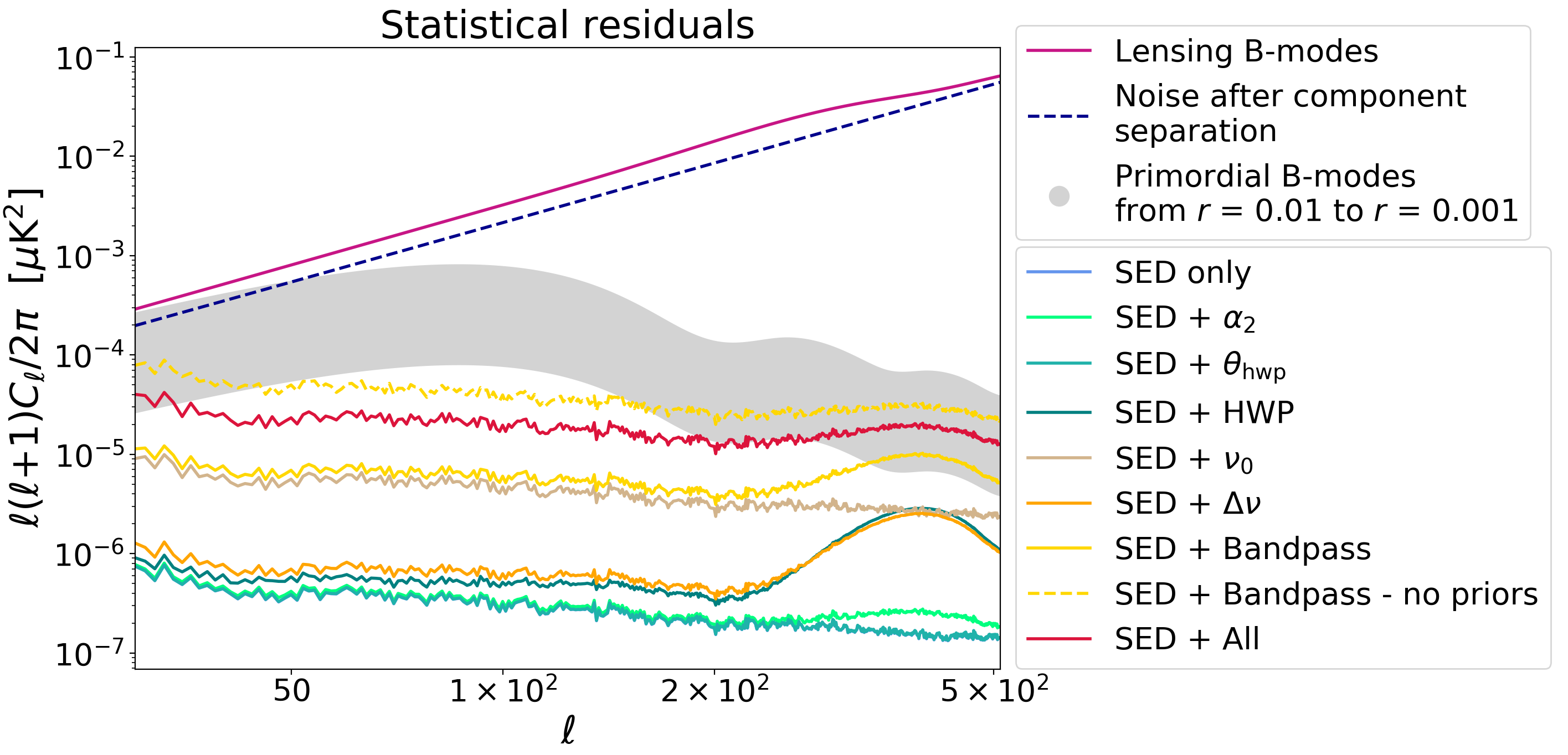}
    \caption{Foreground statistical residuals on B-modes, for various instrumental parameter sets. As expected, statistical residuals scale as statistical uncertainty on the estimated parameters. The bump at high $\ell$ is due to statistical uncertainty dominated by uncertainty on instrumental parameters, resulting into a CMB E- to B-mode leakage.}
     \label{fig:stat_res}
\end{figure*}

At first, we assume the same model underlying the data set as it is used for its analysis. In this case, we expect no systematic residuals, and we can therefore limit ourselves to statistical residuals as defined in Eqs.~(\ref{eq:res_stat}). We compute foreground residuals for all cases studied in the previous section as shown in Figure \ref{fig:stat_res}.

As expected, the level of statistical residuals scales directly as the statistical uncertainties on recovered parameters: the more parameters, the greater the uncertainties and the higher the statistical residuals. Including priors on bandpass parameters lower the level of residuals as expected. 

We note that, despite being very similar in shape at low $\ell$, the various residuals curves differ occasionally at higher $\ell$: the bump is due to the leakage of CMB polarisation signal (E-modes) to the total residuals, referred to as the CMB residuals in Eq.~(\ref{eq:cmb-res}). The relative amplitude of both residual terms (foregrounds and CMB) depends on the instrumental parameters. For the parameters which do not affect much the determination of the spectral parameters, the additional foreground residuals and the CMB residuals scale the same way, as they both are proportional to the uncertainties on these parameters. As the CMB E-mode power dominates that of the foregrounds at small angular scales the CMB residuals may then become dominant at high-$\ell$s. If the instrumental parameters are strongly coupled to the spectral ones, there is an extra increase of the foreground residuals because of the larger errors on the spectral parameters themselves and the foreground residuals may therefore dominate over the entire range of angular scales considered here. In such cases no high-$\ell$ bump is seen in the residuals spectra. 

\begin{figure*}
    \centering
    \includegraphics[width=\textwidth]{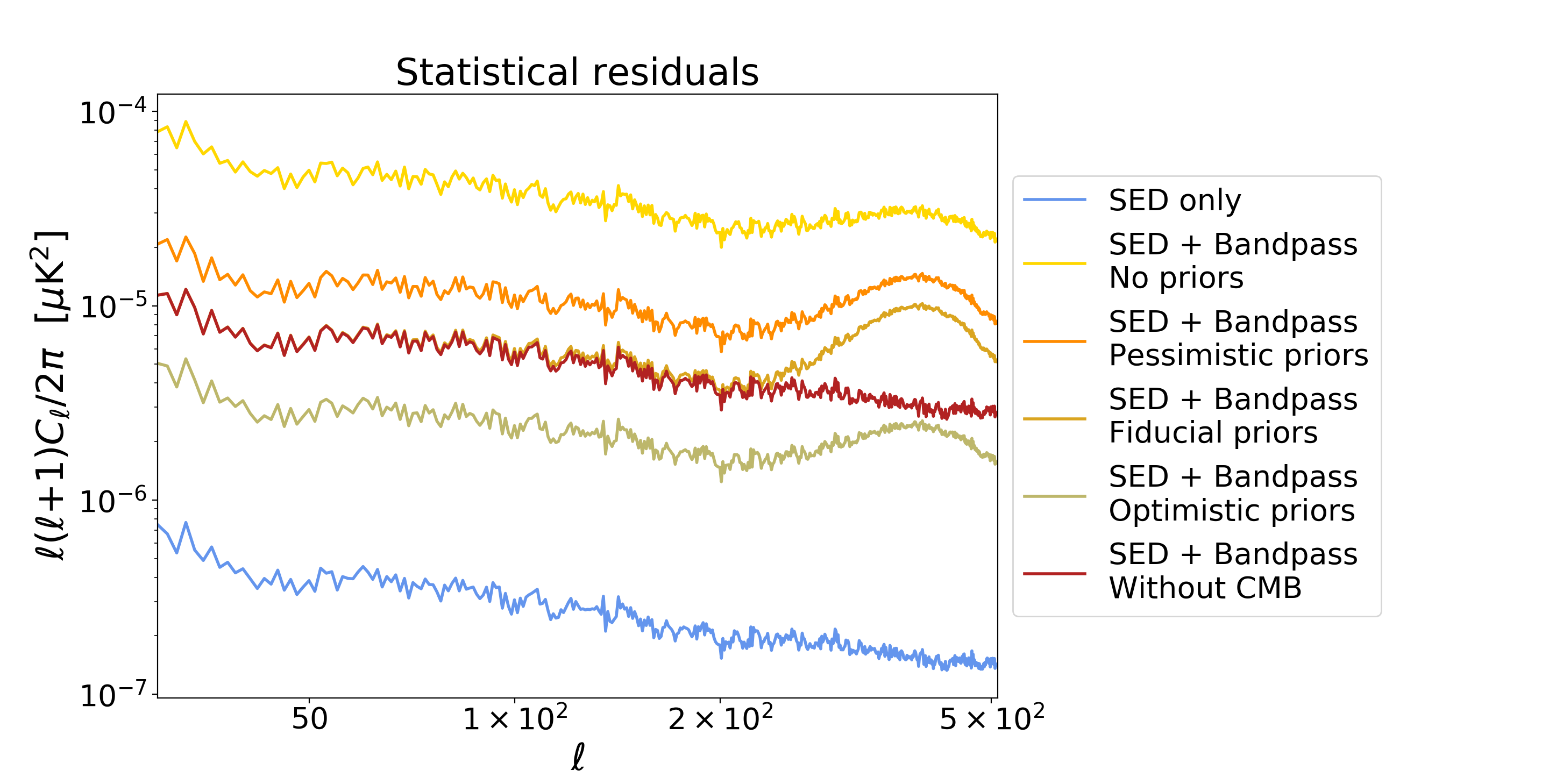}
    \caption{Foreground statistical residuals on B-modes, with and without including CMB contribution. We also compare different level of priors on bandwidths, to show the effective reduction of statistical residuals by choosing more stringent priors. The case without CMB is computed with fiducial priors}
     \label{fig:stat_res_no_cmb}
\end{figure*}

We validate these expectations by computing residuals assuming no CMB and show that this distinctive feature disappears. This is illustrated in Figure~\ref{fig:stat_res_no_cmb}.

As we observed in section~\ref{bp-params} (see Figure~\ref{fig:sigma_beta}), marginalised uncertainties on spectral parameters in the \{SED + $\nu_0$\} case are much higher than in the \{SED only\} case, while there is only a negligible increase in the \{SED + $\Delta \nu$\} cas. This is reflected in Figure~\ref{fig:stat_res}, where at low $\ell$  the statistical residuals in the \{SED + $\nu_0$\} case are higher than in the \{SED + $\Delta \nu$\} case. Moreover, in the former case, unlike in the latter, there is no high--$\ell$ bump in the residuals, and the residuals are dominated by the foreground term. We also note that including either the bandpass widths, \{SED+$\Delta \nu$\}, or the HWP parameters, \{SED+HWP\}, gives rise to residuals which are very much comparable across the entire range of the considered angular scales. This is because in both these cases the spectral parameters uncertainties are not significantly affected by these instrumental parameters, whereas the instrumental parameters themselves are estimated with similar precision. Therefore the levels of the foreground and CMB residuals are expected to be indeed comparable.

In Figure \ref{fig:stat_res_no_cmb} we show the impact of priors on the residuals. We see that more stringent priors indeed lower the overall level of the residuals but also that the relative import of the CMB residual progressively decreases.

\subsubsection{Systematic residuals} 
\label{bp-syst}
\begin{figure*}
    \centering
    \includegraphics[width=\textwidth]{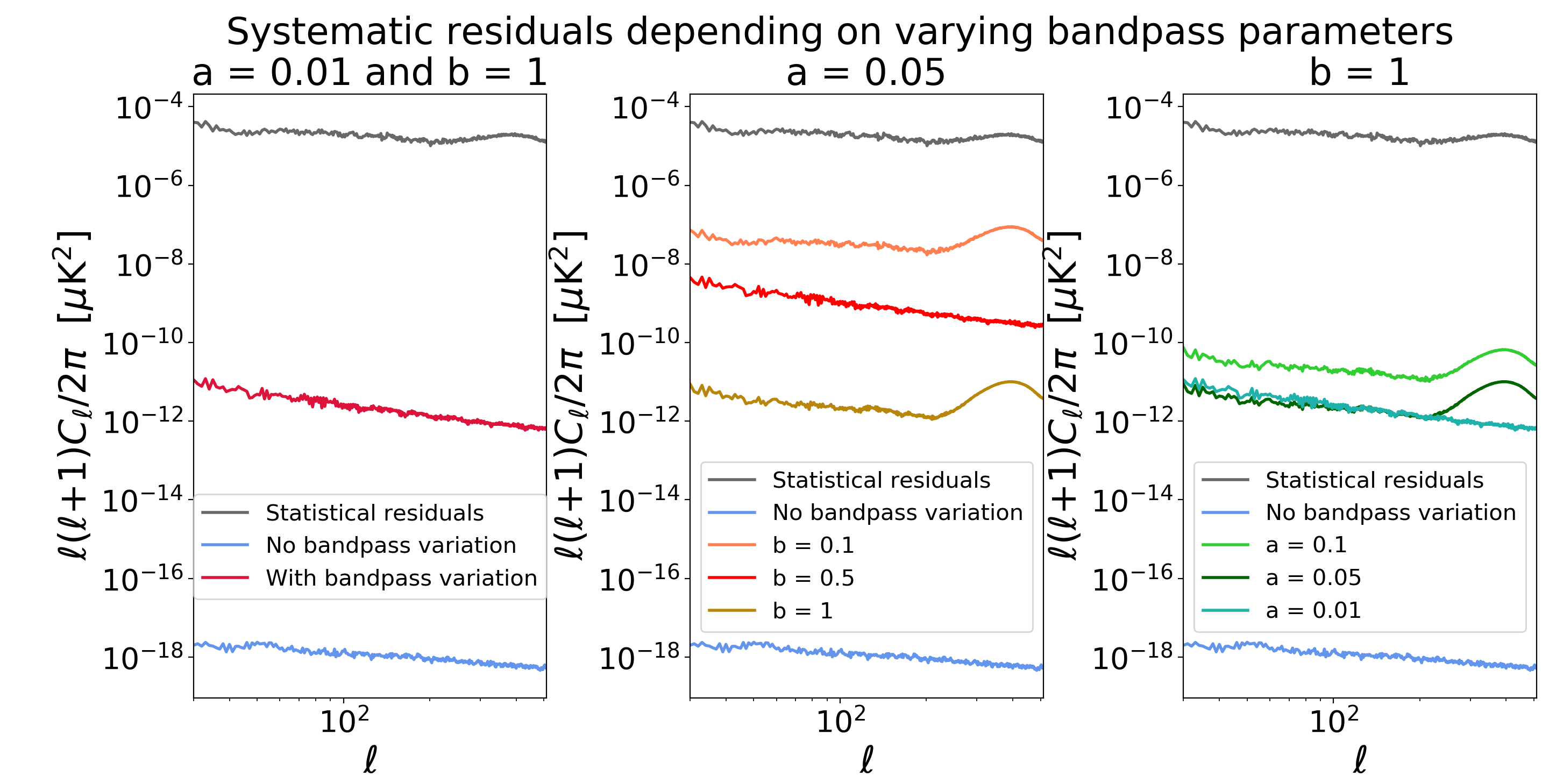}
    \caption{Left: Statistical and systematic residuals, showing the increase of systematic residuals when introducing a mismatch between bandpass in data and model when considering the \{SED + All\} case. Center and right: scaling of systematic residuals depending on bandpass variation, compared to statistical residuals. We consider here all 20 parameters, i.e. the \{SED + All\} case. The systematic residuals in the case with no bandpass variation should be zero, the light blue curve defines therefore numerical precision of our computations.}
    \label{fig:syst_all}
\end{figure*}

We now introduce a mismatch between simulated data and the model. As outlined in section \ref{bp-design}, we investigate the effect of slowly varying, more realistic bandpass in the data, while we keep on using a simple smoothed top-hat model in the analysis. We consider the following cases:
\begin{itemize}
    \item SED only
    \item SED + HWP
    \item SED + Bandpass
    \item SED + All
\end{itemize}
For each case, we compute residuals when introducing a discrepancy between bandpass data and model, with $a = 0.01$ and $b = 1$ in the varying bandpass. We compare them to residuals in the same instrumental configuration, but with no bandpass variation in the input data. For both statistical (Eq.~(\ref{eq:res_stat})) and systematic residuals (Eq.~(\ref{eq:res_syst})), we verify that they behave as expected:
\begin{itemize}
    \item no notable difference between statistical residuals, as they are unaffected by discrepancy between data and model;
    \item increase of systematic residuals as they are precisely sensitive to the mismatch between data and its model.
\end{itemize}
We demonstrate that this is true for all cases and, for reference, show the result for the \{SED + All\} case in the left panel of Figure \ref{fig:syst_all}.

The level of systematic residuals depends on parameters $a$ and $b$ (defined in section \ref{bp-design}), which set the amplitude and frequency of bandpass variations. 
We investigate this dependence focusing on the \{SED + All\} case, as this is the case where we expect the biggest impact. We consider various \{$a,b$\} combinations and compute corresponding residuals. We find that, for a fixed value of $a$, i.e., the amplitude of the bandpass variation, a smaller $b$, corresponding to variations on larger scales, leads to an increase in the systematic residuals. In contrary, the effects of small-scale features in the bandpass are quickly becoming very small, suggesting that characterising the bandpasses with a $\sim$ $1$GHz sampling as this is most often the case~\cite{pb-fts} in practise, should indeed be sufficient. Moreover, for a given value of $b$, the systematics residuals increase with increasing value of $a$. This is an expected effect, since $a$ sets the amplitude of bandpass variation. These two effects are demonstrated on Figure \ref{fig:syst_all}. 
In section \ref{r-bp} hereafter, we comment on the impact of the systematic residuals on $r$ estimation.

\subsection{Constraints on tensor-to-scalar ratio}
\label{r}

With the residuals estimated for various cases, we now estimate their impact on the determination of the tensor-to-scalar ratio $r$. We use the cosmological likelihood in Eq.~(\ref{eq:cosmo-likelihood}) and consider two models for the assumed covariance of the CMB signal, $\mathbf{C}$, as described in section~\ref{cosmo-likelihood}. These correspond to two extreme cases when we either assume complete ignorance of the residuals present in the cleaned CMB maps, Eq.~(\ref{eq:no_deprojection}), or when the statistical residuals are modeled and deprojected consistently, Eq.~(\ref{eq:deprojectCov}). We focus only on the B-mode polarization and the CMB covariance includes only the BB power spectra parametrised by the tensor-to-scalar ratio $r$ as elaborated on in section~\ref{sec:mapParams}.

\subsubsection{Cases with statistical residuals only} 
In this case the model underlying the data coincides with that assumed in the analysis and the only residuals are statistical. We first assume that the actual value of $r$ is zero.

\begin{figure*}
    \centering
    \includegraphics[width=0.9\textwidth]{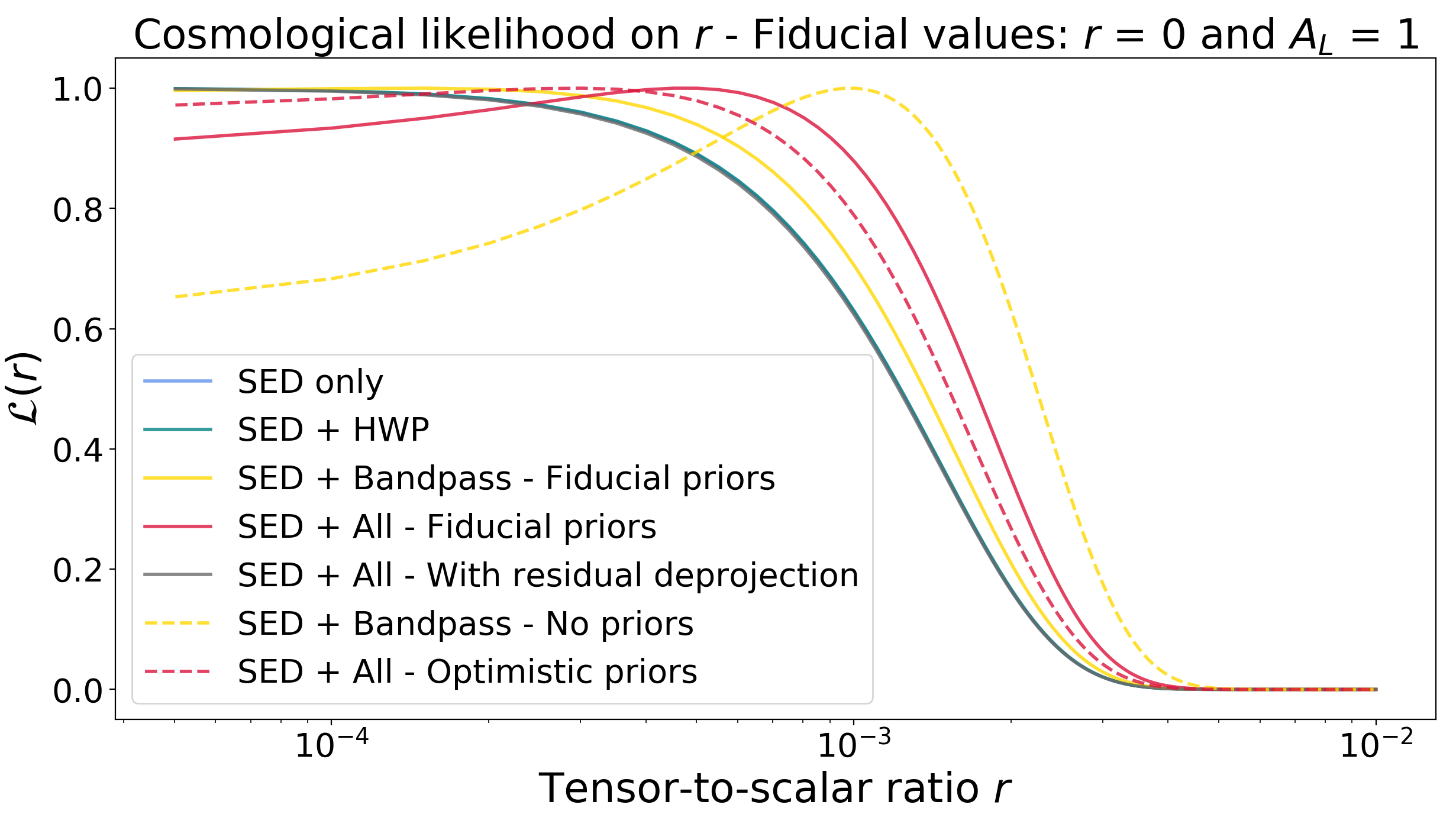}
    \caption{Cosmological likelihood Eq.~(\ref{eq:cosmo-likelihood}) as a function of the tensor-to-scalar ratio $r$, with its fiducial value assumed to be $0$. In all cases, but one as marked in the legend, no deprojection is applied. In these cases the priors are shown to be efficient to alleviate the bias on $r$. The deprojection of the residuals not only resolves the issue of the bias (due to the statistical residuals) but also recovers the lowest, theoretically possible uncertainty on $r$.}
    \label{fig:r}
\end{figure*}

In this first approach we ignore the residuals present in the recovered CMB map and therefore expect that any residual, be it statistical or systematic, will result in a bias on $r$ which in turn will depend on their level. The results are shown in Figure \ref{fig:r} and the corresponding values of $r$ and $\sigma(r)$ for all instrumental parameter sets are collected in Table~\ref{tab:r_no_deproj}.
\begin{table}
    \centering
    \begin{tabular}{|c|c|c|}
    \hline 
         & $r (\times 10^{-5}$) &  $\sigma(r)  (\times 10^{-3})$ \\
    \hline
        SED only & 0.819 & 1.47\\
    \hline
        SED + $\alpha_2$ & 0.870 & 1.47\\
    \hline
        SED + $\theta_\mathrm{hwp}$ & 0.823 & 1.47\\
    \hline
        SED + HWP & 1.15 & 1.47\\
    \hline
        SED + $\nu_0$ - fiducial priors & 8.40 & 1.47\\ 
    \hline
        SED + $\Delta \nu$ - fiducial priors & 1.38 & 1.47 \\ 
    \hline
        SED + Bandpass - no priors & 98.6 & 1.51\\
    \hline
        SED + Bandpass - fiducial priors & 14.1 & 1.48 \\
    \hline
        SED + All - fiducial priors & 47.6 & 1.49 \\
    \hline
        SED + All - optimistic priors & 29.1 & 1.48  \\
    \hline
    \end{tabular}
    \caption{Best-fit values for $r$ and $\sigma(r)$ without the residuals deprojection. Corresponding cosmological likelihoods for selected cases are shown in Figure \ref{fig:r}. The recovered uncertainty for the \{SED only\} case is compatible with the Simons Observatory forecasted performance~\cite{SO}.}
    \label{tab:r_no_deproj}
\end{table}

As expected, the higher are the residuals, the bigger is the bias on $r$. We can therefore suppress the bias by suppressing the residuals. This can be achieved with help of additional priors on instrumental parameters. As shown in Figure~\ref{fig:r} and Table \ref{tab:r_no_deproj}, this way is indeed efficient in reducing the bias on $r$ and ensuring that it does not lead to an erroneous but statistically significant detection. Its downside is that it may call for high precision priors. This can be particularly demanding if a large number of relevant instrumental effects is present.  

The statistical residuals can be effectively margi\-nali\-sed, or deprojected, by using the model covariance in Eq.~(\ref{eq:deprojectCov}), therefore minimising potential bias of the measured value of $r$. In this case, we may hope that the need for high precision priors can at least be partly alleviated. This expectation is indeed confirmed in the cases studied here, as it is shown with a gray solid line in Figure~\ref{fig:r}. The loss of accuracy and precision seems to be in this case negligible. In more realistic cases and in the presence of a potentially significant mismatch between the assumed sky and instrument models, and the true ones, the deprojection may not be however as successful as in the simple cases studied here. We expect therefore that the calibration priors will continue playing a key role in the component separation process either in order to suppress the residuals, as in the first method described above, or to demonstrate that the deprojection was indeed performed successfully.

\begin{figure}
    \centering
    \includegraphics[width=0.45\textwidth]{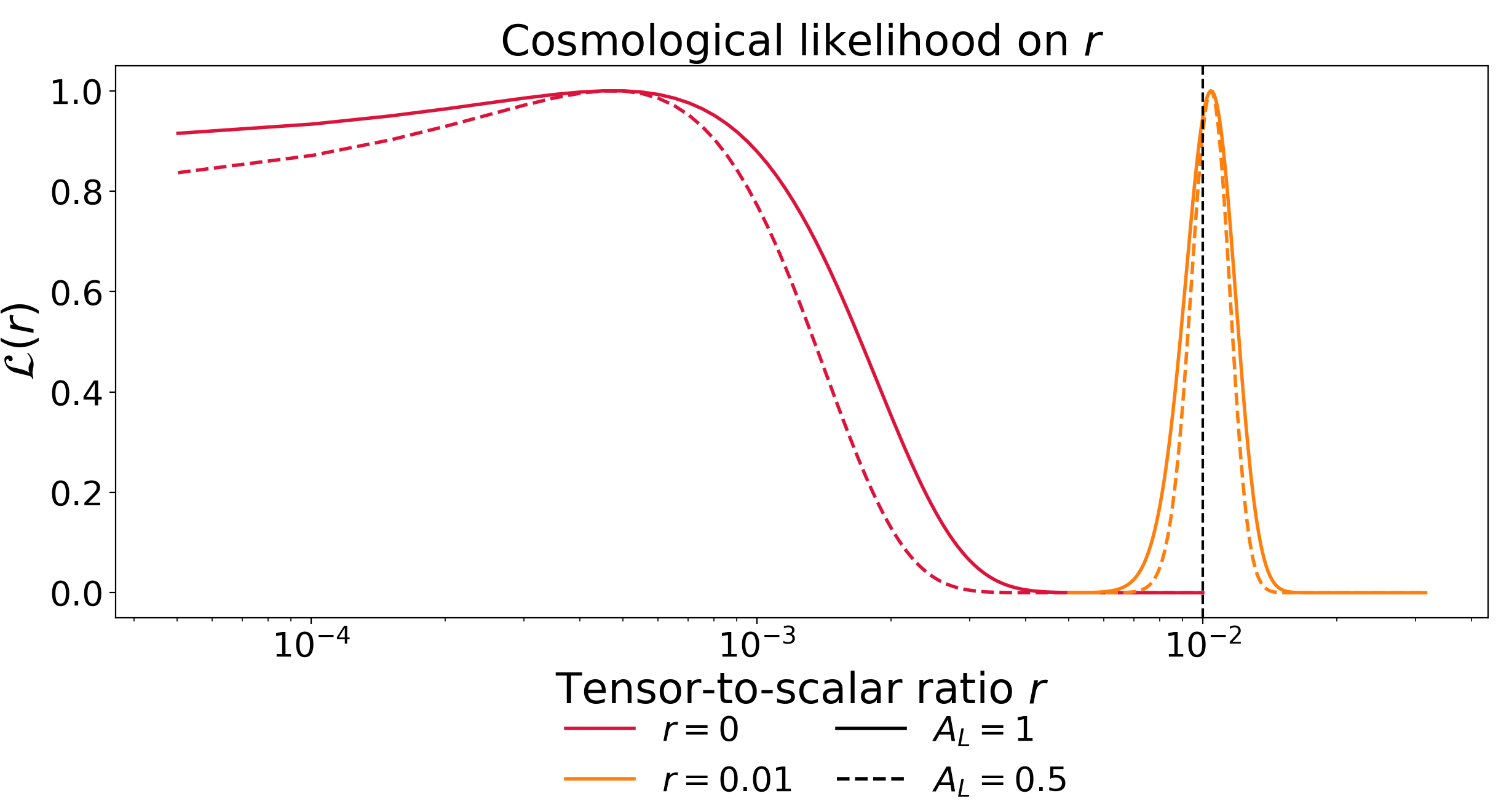}
    \caption{Cosmological likelihood on $r$ without the residual deprojection for two different fiducial values of the tensor-to-scalar ratio, $r=0$ or $r=0.01$, and assuming either no, $A_L=1$, or partial, $50$\% delensing, $A_L=0.5$. The shown results are for the case of \{SED + All\}, thus including priors on bandpass parameters.}
    \label{fig:various-r}
\end{figure}

We also test our framework with other values of cosmological parameters considering for the tensor-to-scalar ratio a value of $r = 0.01$ consistent with what typical Stage 3 experiment plan to achieve and assuming partial delensing with $A_L = 0.5$ in addition to the no-delensing case $A_L=1$. We limit ourselves to the case with all instrumental parameters and include priors, \{SED + All\}. Results without the residual deprojection are shown in Figure \ref{fig:various-r}. We show that in all considered cases, our estimates of the value of $r$ are consistent with the true values with precision much better than the $1{\rm -}\sigma(r)$ uncertainty, even if we do not deproject statistical residuals. 

\subsubsection{Impact of bandpass imperfections}
\label{r-bp}
As we have shown previously, the mismatch in the bandpass parameters unavoidably leads to the presence of the systematic residuals. These were found to be sub-dominant with respect to the statistical residuals in the all the cases studied in this work, section \ref{bp-syst}. As long as no deprojection is applied both types of the residuals impact the estimates of $r$ in the same way. Consequently, in all our cases the effect on $r$ due to the systematic residuals is found to be subdominant and negligible. This remains to be the case also when the statistical residuals are deprojected. The impact of the systematic residuals on the value of $r$ is then very minor and much smaller than the statistical error on $r$. This is due to the low absolute level of the systematic residuals present in the studied cases.

\section{Conclusions}
\label{conclusions}

In this work we have extended the standard CMB data analysis pipeline to include explicitly a treatment of instrumental effects. We focused on two key data analysis stages, map-making and component separation, and considered, as an example, instrumental effects related to the presence of a broadband HWP in the instrument optical design, and bandpasses defining the frequency bands of the observations.

We have subsequently implemented the proposed framework as part of the performance forecasting tool {\sc xForecast}~\cite{xforecast,xForecast2}, extending it to account for instrumental effects. This has allowed us to propagate the impact of the instrumental effects all the way to cosmological constraints. 

We have applied the method in the context of a modern CMB experiment of the 3rd generation, modeled on the small aperture telescopes of Simons Observatory~\cite{SO} assessing its performance in the light of setting constraints on the tensor-to-scalar ratio $r$. We have discussed the role and impact of the instrumental parameters on the cosmological constraints, see also \cite{max} for complementary analysis specific to the Simons Observatory. We have shown how calibration information can be incorporated in the analysis and how to determine which instrumental parameters may need such external prior information in order to not compromise the analysis results. We have demonstrated that in the studied cases with help of either suitable priors or data analysis techniques we can efficiently suppress the systematic biases and control the statistical uncertainties. In turn, these results can provide insights about the precision level for calibration of various instrumental parameters required given pre-defined science goals.

The cases studied here were clearly over-idealized. This concerns both the models assumed for the foregrounds and the instruments. This is because our main purpose was to describe, validate, and demonstrate the proposed approach. However, more complex foreground models can be included following the procedures of~\cite{xForecast2}, as can more realistic instrument models. As an example we have studied the case of the bandpasses where in addition we allowed for a mismatch between the underlying bandpass model and the one used for the analysis. We leave a  more thorough and exhaustive exploration of other possible effects to future work.

The key feature of the proposed approach is the assumption that efficient, parametric models of the instrumental effects can be devised and then used to mitigate their impact. We emphasize that such models may be merely approximate and phenomenological, as in our bandpass example. The proposed forecasting tool allows to evaluate efficiency (and sufficiency) of a model given more complex real instruments which could be characterized by non-parametric models derived from actual measurements.

\section*{Acknowledgements}

We acknowledge the use of \texttt{healpy}~\cite{healpy}, \texttt{PySM}~\footnote{\url{https://pysm3.readthedocs.io/en/latest/#}} and \texttt{fgbuster}~\footnote{\url{https://fgbuster.github.io/fgbuster/index.html}} software packages.
We thank Maximilian Abitbol, David Alonso and Davide Poletti for useful comments and discussions. 

This work was supported by the French National Research Agency (ANR) grants, ANR-B3DCMB, (ANR-17-CE23-0002), and ANR-BxB (ANR-17-CE31-0022).

\FloatBarrier

\begin{widetext}
\appendix
\section{HWP optics}
\label{hwp-appendix}

We give here the detailed expressions of $\mathbf{C}_{0\mathrm{i};\, \mathrm{k}}$ and $\mathbf{S}_{0\mathrm{i};\, \mathrm{k}}$ coefficients introduced in Eq.~(\ref{eq:datamodel-simple}, as a function of HWP Mueller matrix elements $\boldsymbol{\mu}_{ij}$. We note that $\boldsymbol{\mu}_{0\mathrm{i}}$ and $\boldsymbol{\mu}_{\mathrm{i}0}$ elements are zeros by design of the HWP, so we have:

\begin{align}
\label{eq:wobble-coeff}
\mathbf{C}_{01;\, 0}(\nu) =  &\frac{1}{2}\,(\boldsymbol{\mu}_{11}+\boldsymbol{\mu}_{22})\cos{(2\eta_\nu)}\\
\mathbf{C}_{01;\, 4}(\nu) =  &\frac{1}{2}[\,(\boldsymbol{\mu}_{11}-\boldsymbol{\mu}_{22})\cos{(2\eta_\nu)} - (\boldsymbol{\mu}_{12}+\boldsymbol{\mu}_{21})\sin{(2\eta_\nu)}] \\
\mathbf{S}_{01;\, 0}(\nu) = & - \frac{1}{2}\,(\boldsymbol{\mu}_{11}+\boldsymbol{\mu}_{22})\sin{(2\eta_\nu)}\\
\mathbf{S}_{01;\, 4}(\nu)  = &  \frac{1}{2}[\,(\boldsymbol{\mu}_{12}+\boldsymbol{\mu}_{21})\cos({2\eta_\nu}) + (\boldsymbol{\mu}_{11}-\boldsymbol{\mu}_{22})\sin{(2\eta_\nu)}]\\
\mathbf{C}_{02;\, 0}(\nu)  = & - \frac{1}{2}\,(\boldsymbol{\mu}_{11}+\boldsymbol{\mu}_{22})\sin{(2\eta_\nu)}\\
\mathbf{C}_{02;\, 4}(\nu) = &  \frac{1}{2}[\,(\boldsymbol{\mu}_{12}+\boldsymbol{\mu}_{21})\cos{(2\eta_\nu)}  +(\boldsymbol{\mu}_{11}-\boldsymbol{\mu}_{22})\sin{(2\eta_\nu)}]\\\
\mathbf{S}_{02;\, 0}(\nu) = &  - \frac{1}{2}\,(\boldsymbol{\mu}_{11}+\boldsymbol{\mu}_{22})\cos{(2\eta_\nu)}\\
\label{eq:wobble-coeff-end}
\mathbf{S}_{02;\, 4}(\nu) = &  -\frac{1}{2}[\,(\boldsymbol{\mu}_{11}-\boldsymbol{\mu}_{22})\cos{(2\eta_\nu)}  -(\boldsymbol{\mu}_{12}+\boldsymbol{\mu}_{21})\sin{(2\eta_\nu)}]
\end{align}

We have explicitly included the frequency dependence of the coefficients, as Mueller matrix elements of the HWP (through $\delta$ defined in Eq.~(\ref{eq:delta})) and sinuous antenna depend on observing frequency. This allows to naturally take into account the instrumental frequency-dependent effects in the model.

\section{Coefficients in the 3-layer HWP case}
\label{3layer-appendix}

In Eqs.~(\ref{eq:wobble-coeff} - \ref{eq:wobble-coeff-end}) in Appendix \ref{hwp-appendix}, we expressed the coefficients of the full optics chain Mueller matrix, as a function of HWP Mueller matrix coefficients $\boldsymbol{\mu}_{ij}$. These coefficients can be computed from Eq.~(\ref{eq:multi-layer-hwp}) for any HWP configuration, or obtained from measurements. 

We give here the full analytical expression of these coefficients in the instrumental configuration that we detailed in section \ref{hwp-config}. We recall that we have two parameters per HWP:

\begin{itemize}
    \item $\alpha_2$: central layer rotation angle;
    \item $\theta_\mathrm{hwp}$: thickness of one layer.
\end{itemize}

For clarity reasons, we use $\delta$ instead of $\theta_\mathrm{hwp}$, defined as in Eq.~(\ref{eq:delta}):

\begin{equation}
    \delta = \frac{2\pi\theta_\mathrm{hwp}\,|n_o-n_e| \nu }{c}.
\end{equation}

We can then express HWP Mueller matrix coefficients as:

\begin{align}
    \boldsymbol{\mu}_{11} & = \cos^2(2  \alpha_2) \, + \, \cos(\delta) \, \sin^2(2 \alpha_2) \nonumber \\
    \boldsymbol{\mu}_{12} & = \boldsymbol{\mu}_{21} = \sin(2 \alpha_2) \, \cos(2\alpha_2) \left[ \cos^2(\delta)-\cos(\delta) \right]  - \, \sin^2(\delta) \, \sin(2\alpha_2) \nonumber \\
     \boldsymbol{\mu}_{22} & = \cos^2(\delta) \, \sin^2(2\alpha_2) \, + \, \cos^3(\delta) \, \cos^2(2\alpha_2) \nonumber \\
   & - \left[ 2 \sin^2(\delta) \, \cos(\delta) \, \cos(2\alpha_2) + \, \sin^2(\delta) \, \cos(\delta) \right]
\end{align}

\section{Cosmological likelihood}
 \label{deprojection}
Our implementation of the cosmological likelihood is based on the one proposed in Appendix C of \cite{xforecast}. The cosmological likelihood Eq.~(\ref{eq:cosmo-likelihood}) is split into three terms:

\begin{equation}
     \langle S^\mathrm{cos} \rangle = \tr \mathbf{C}^{-1} \mathbf{\hat{C}} \, + \, \tr \mathbf{C}^{-1}(\mathbf{E} - \mathbf{\hat{C}}) \, + \, \ln \det \mathbf{C},
\end{equation}
where $\mathbf{E}$ is the true (observed) signal covariance matrix. Note that when there are no systematic residuals (our main case in this work, except when we consider bandpass variation), the second term of the sum actually vanishes. As outlined in section \ref{cosmo-likelihood}, we consider two cases for the assumed signal covariance matrix: the no deprojection case, where $\mathbf{C} = \mathbf{C}^\mathrm{cmb}$; and the deprojection case, where $\mathbf{C} = \mathbf{C}^\mathrm{cmb} + \mathbf{C}^\mathrm{stat.}$.

\subsection{No deprojection case}
In the no deprojection case, the three terms of the cosmological likelihood are written as:

\begin{align}
    \tr \mathbf{C}^{-1} \mathbf{\hat{C}} & = \sum_\ell \left[\frac{(2 \ell +1)}{C_\ell} \left(\hat{C}_\ell \, + \tr [\mathbf{\Sigma} \otimes_\ell(\mathbf{\Tilde{Y}}^{(1)}, \mathbf{\Tilde{Y}}^{(1)})] \right)\right] \\ 
    \tr \mathbf{C}^{-1}(\mathbf{E} - \mathbf{\hat{C}}) & = \sum_\ell \left[\frac{(2\ell +1)}{C_\ell} \left(\otimes_\ell (\mathbf{\Tilde{y}},\mathbf{\Tilde{y}}) \, + \, \otimes_\ell (\mathbf{\Tilde{z}},\mathbf{\Tilde{y}})  \, + \, \otimes_\ell (\mathbf{\Tilde{y}},\mathbf{\Tilde{z}}) \right) \right] \\
    \ln \det \mathbf{C} & = \ln \det \mathbf{C}^\mathrm{cmb}
\end{align}

\subsection{Deprojection case}
In the deprojection case, we consider that we have a model for the statistical residuals, that we include in the modelled covariance matrix $\mathbf{C}$. The three terms of the cosmological likelihood then reads as (note that this case corresponds to previous implementation of the formalism, and thus the following equations are exactly equations (C9), (C10) and (C12) of \cite{xforecast}):

\begin{align}
    \tr \mathbf{C}^{-1} \mathbf{\hat{C}} & = \sum_\ell \left[(2 \ell +1)\frac{\hat{C}_\ell}{C_\ell} \left(1 - C_\ell^{-1} \tr[\mathbf{U} \otimes_\ell (\mathbf{\Tilde{Y}}^{(1)}, \mathbf{\Tilde{Y}}^{(1)})] \right) + \frac{(2 \ell +1)}{C_\ell} \tr [\mathbf{\Sigma} \otimes_\ell(\mathbf{\Tilde{Y}}^{(1)}, \mathbf{\Tilde{Y}}^{(1)})]\right] \nonumber \\
    & - \, \sum_{\ell,\ell'} \frac{(2 \ell +1)}{C_\ell} \frac{(2 \ell' +1)}{C_\ell'} \tr[\mathbf{U} \otimes_{\ell'}(\mathbf{\Tilde{Y}}^{(1)}, \mathbf{\Tilde{Y}}^{(1)})\mathbf{\Sigma}\otimes_{\ell}(\mathbf{\Tilde{Y}}^{(1)}, \mathbf{\Tilde{Y}}^{(1)})] \\ 
    \tr \mathbf{C}^{-1}(\mathbf{E} - \mathbf{\hat{C}}) & = \sum_\ell \left[\frac{2\ell +1}{C_\ell} \left(\otimes_\ell (\mathbf{\Tilde{y}},\mathbf{\Tilde{y}}) \, + \, \otimes_\ell (\mathbf{\Tilde{z}},\mathbf{\Tilde{y}})  \, + \, \otimes_\ell (\mathbf{\Tilde{y}},\mathbf{\Tilde{z}}) \right) \right] \nonumber \\
     -  \sum_{\ell,\ell'}\frac{(2 \ell +1)}{C_\ell} \frac{(2 \ell' +1)}{C_\ell'} & \tr\left[\mathbf{U}\left(\otimes_{\ell'} (\mathbf{\Tilde{Y}}^{(1)},\mathbf{\Tilde{y}})\otimes_{\ell} (\mathbf{\Tilde{y}},\mathbf{\Tilde{Y}}^{(1)}) + \otimes_{\ell'} (\mathbf{\Tilde{Y}}^{(1)},\mathbf{\Tilde{y}})\otimes_{\ell} (\mathbf{\Tilde{z}},\mathbf{\Tilde{Y}^{(1)}}) + \otimes_{\ell'} (\mathbf{\Tilde{Y}}^{(1)},\mathbf{\Tilde{z}})\otimes_{\ell} (\mathbf{\Tilde{y}},\mathbf{\Tilde{Y}^{(1)}}) \right)\right]\\
    \ln \det \mathbf{C} & = \ln \frac{\det \mathbf{C}^\mathrm{cmb}}{\det \mathbf{U}},
\end{align}
where we have introduced $\mathbf{U}$ as defined in Eq.~(C1) in \cite{xforecast}:

\begin{equation}
    \mathbf{U} \equiv \left(\mathbf{\Sigma}^{-1} +\mathbf{\Tilde{Y}}^{(1)\dagger}\mathbf{C}^\mathrm{cmb}\mathbf{\Tilde{Y}}^{(1)}\right)^{-1}
\end{equation}
\end{widetext}

\bibliography{references}

\end{document}